\documentclass[twocolumn,tighten]{aastex63}
\pdfoutput=1
\usepackage{graphicx}
\usepackage{ulem}
\usepackage{amsmath}
\usepackage{wrapfig}
\usepackage{appendix}
\usepackage{enumitem}
\usepackage{multirow}
\usepackage{floatrow}
\usepackage{color}
\usepackage{longtable}

\newcommand\msun{\, \rm M_\odot}

\newcommand\zsun{\, \rm Z_\odot}
\newcommand\kms{\, \rm km\,s^{-1}}

\begin{document}
\shorttitle{Demographics of triple systems in dense star clusters}
\shortauthors{Fragione et al.}

\title{Demographics of triple systems in dense star clusters}

\author[0000-0002-7330-027X]{Giacomo Fragione}
\affil{ Department of Physics \& Astronomy, Northwestern University, Evanston, IL 60208, USA}
\affil{Center for Interdisciplinary Exploration \& Research in Astrophysics (CIERA), Northwestern University, Evanston, IL 60208, USA}
\email{giacomo.fragione@northwestern.edu}

\author[0000-0001-5285-4735]{Miguel A. S. Martinez}
\affil{ Department of Physics \& Astronomy, Northwestern University, Evanston, IL 60208, USA}
\affil{Center for Interdisciplinary Exploration \& Research in Astrophysics (CIERA), Northwestern University, Evanston, IL 60208, USA}

\author[0000-0002-4086-3180]{Kyle Kremer}
\affil{ Department of Physics \& Astronomy, Northwestern University, Evanston, IL 60208, USA}
\affil{Center for Interdisciplinary Exploration \& Research in Astrophysics (CIERA), Northwestern University, Evanston, IL 60208, USA}

\author[0000-0002-3680-2684]{Sourav Chatterjee}
\affil{Department of Astronomy \& Astrophysics, Tata Institute of Fundamental Research, Homi Bhabha Road, Navy Nagar, Colaba, Mumbai 400005, India}

\author[0000-0003-4175-8881]{Carl L. Rodriguez}
\affil{Astronomy Department, Harvard University, 60 Garden St., Cambridge, MA 02138, USA}

\author[0000-0001-9582-881X]{Claire S. Ye}
\affil{ Department of Physics \& Astronomy, Northwestern University, Evanston, IL 60208, USA}
\affil{Center for Interdisciplinary Exploration \& Research in Astrophysics (CIERA), Northwestern University, Evanston, IL 60208, USA}

\author[0000-0002-9660-9085]{Newlin C. Weatherford}
\affil{ Department of Physics \& Astronomy, Northwestern University, Evanston, IL 60208, USA}
\affil{Center for Interdisciplinary Exploration \& Research in Astrophysics (CIERA), Northwestern University, Evanston, IL 60208, USA}

\author[0000-0002-9802-9279]{Smadar Naoz}
\affil{Department of Physics and Astronomy, University of California, Los Angeles, CA 90095, USA}
\affil{Mani L. Bhaumik Institute for Theoretical Physics, UCLA, Los Angeles, CA 90095, USA}

\author[0000-0002-7132-418X]{Frederic A. Rasio}
\affil{ Department of Physics \& Astronomy, Northwestern University, Evanston, IL 60208, USA}
\affil{Center for Interdisciplinary Exploration \& Research in Astrophysics (CIERA), Northwestern University, Evanston, IL 60208, USA}

\begin{abstract}
Depending on the stellar type, more than $\sim 50$\% and $\sim 15\%$ of stars in the field have at least one and two stellar companions, respectively. Hierarchical systems can be assembled dynamically in dense star clusters, as a result of few-body encounters among stars and/or compact remnants in the cluster core. In this paper, we present the demographics of stellar and compact-object triples formed via binary--binary encounters in the \texttt{CMC Cluster Catalog}, a suite of cluster simulations with present-day properties representative of the globular clusters (GCs) observed in the Milky Way. We show how the initial properties of the host cluster set the typical orbital parameters and formation times of the formed triples. We find that a cluster typically assembles hundreds of triples with at least one black hole (BH) in the inner binary, while only clusters with sufficiently small virial radii are efficient in producing triples with no BHs, as a result of the BH-burning process. We show that a typical GC is expected to host tens of triples with at least one luminous component at present day. We discuss how the Lidov-Kozai mechanism can drive the inner binary of the formed triples to high eccentricities, whenever it takes place before the triple is dynamically reprocessed by encountering another cluster member. Some of these systems can reach sufficiently large eccentricities to form a variety of transients and sources, such as blue stragglers, X-ray binaries, Type Ia Supernovae, Thorne-Zytkow objects, and LIGO/Virgo sources.
\vspace{1cm}
\end{abstract}

\section{Introduction}
\label{sect:intro}

Stellar multiplicity is an omnipresent outcome of the star-formation process \citep{duc13}. More than $\sim 50$\% of stars are thought to have at least one stellar companion \citep[e.g.,][]{tok14a}. \citet{tok14b} showed that at least $\sim 13$\% of F-type and G-type dwarf stars in the Hipparcos sample live in triple systems (an inner binary orbited by an outer companion), while \citet{rid15} found a relatively large abundance of $2$+$2$ quadruples (a binary where the components are themselves binaries) with Robo-AO, the first robotic adaptive optics instrument. \citet{san14} estimated that $\sim 80\%$ of O-type stars have at least one companion and $\sim 25$\% have at least two such companions in their sample. Using a large high-resolution radial velocity spectroscopic survey of B-type and O-type stars, \citet{chi12} estimated that at least $50$-$80$\% of them are multiples. Recently, a black hole of $\sim 5\msun$ has been claimed to live in the triple system HR 6819, $\sim 300$ pc from the Sun \citep{rivinius2020}.

In dense star clusters, hierarchical systems of stars and/or compact remnants can form through few-body (particularly binary--binary) encounters in the clusters' dense cores \citep[e.g.,][]{Fregeau2004,leighgeller2013,antogn2016,frag2019}. In this process, one of the two binaries captures a component of the second binary, with the remaining object leaving the system. \citet{leigh2016} estimated that the branching ratio of this process can be as high as $\sim 10\%$, assuming all equal masses. Therefore, the following questions arise naturally: What is the role of dense star clusters, such as globular clusters (GCs), in dynamically assembling triple systems? What are the properties of these triples? How does this process depend on cluster properties, such as mass, concentration, and metallicity?

GCs represent the ideal environment to study the importance of gravitational dynamics in dense stellar systems and how dynamics shape both cluster evolution and survival \citep[see, e.g.,][]{HeggieHut2003}. Importantly, frequent dynamical encounters between cluster members are fundamental in creating and explaining the existence of a number of exotic populations, such X-ray binaries \citep[e.g.,][]{Clark1975,Verbunt1984,Heinke2005,Ivanova2013,Giesler2018,Kremer2018a}, radio sources \citep[e.g.,][]{Lyne1987,Sigurdsson1995,Ransom2008,Ivanova2008,Fragione2018a,Ye2019}, and gravitational wave (GW) binaries \citep[e.g.,][]{Moody2009,Banerjee2010,Rodriguez2015a,Rodriguez2016a,Askar2017,Banerjee2017,Chatterjee2017a,Chatterjee2017b,Hong2018,Fragione2018b,Samsing2018a,Rodriguez2018b,Zevin2018,Kremer2019b}. However, with the possible exception of \citet{antoninietal2016}, there have been no comprehensive studies about the origin of hierarchical systems in dense star clusters, and how this depends on clusters' primordial properties.

What makes hierarchical triple and multiple systems of particular interest is that they can produce exotic objects, transients, and GW sources over a larger portion of the parameter space compared to binaries. This additional portion is enabled by the Lidov-Kozai (LK) mechanism \citep{lid62,koz62}. In recent years, a number of authors have shown how hierarchical triples are efficient in producing GW sources \citep[e.g.,][]{petrov2017,hamers2018,hoang2018,Fragkoc2019,liu2019,step2019,fragk2020}, tidal disruption events \citep[e.g.,][]{chen2009,fragleigh2018,fragetal2019}, white dwarf (WD) mergers \citep[e.g.,][]{toon2018,fragmet2019}, and millisecond pulsars \citep[e.g.,][]{ford2000}. In this framework, the eccentricity of the inner binary is not constant, but rather oscillates between a minimum and a maximum value (determined by the triple initial configuration), due to the tidal force imposed by the third companion \citep[for a review see][]{naoz2016}. As a result, the inner binary components may be efficiently driven to sufficiently small separations to merge either due to physical collision or dissipation of GWs.

In this paper, we study the role of dense star clusters in producing triple systems of all possible component configurations. We use a grid of $148$ independent cluster simulations \citep[presented in][]{kremer2020}\footnote{\url{https://cmc.ciera.northwestern.edu}}, run using \texttt{CMC} (for \texttt{Cluster Monte Carlo}), which covers roughly the complete range of GCs observed at present day in the Milky Way. We systematically explore the effect of initial virial radii (and subsequent BH dynamics) on clusters of various masses, metallicities, and locations within the Galactic tidal field. We dissect the origin of triples assembled in dense star clusters as a function of the clusters' initial properties, describe the triple demographics, and determine if they can produce exotica, transients, and GW sources.

The paper is organized as follows. In Section~\ref{sect:method}, we describe the numerical method used to evolve our cluster models. In Section~\ref{sect:triporigin}, we analyze the origin of triple systems in star clusters, while in Section~\ref{sect:demogr} we describe their demographics and general properties. In Section~\ref{sect:mergers}, we estimate the transient and GW phenomena as a result of the LK mechanism. Finally, in Section~\ref{sect:conc}, we discuss the implications of our findings and lay out our conclusions.

\section{Methods}
\label{sect:method}

Here, we summarize the methods we use to evolve our population of clusters. For a detailed description see \citet{kremer2020}.

We use \texttt{CMC}, a H\'{e}non-type Monte Carlo code \citep{Henon1971a,Henon1971b,Joshi2000,Joshi2001,Fregeau2003,Chatterjee2010,Chatterjee2013,Pattabiraman2013,Rodriguez2015a}. \texttt{CMC} incorporates the physics relevant to both the overall evolution of the cluster properties and the specific evolution of the stars and compact objects therein.

The main process that shapes the evolution of global properties of clusters is two-body relaxation \citep[e.g.,][]{HeggieHut2003}. In \texttt{CMC}, this is implemented by using the H\'{e}non orbit-averaged Monte Carlo method \citep{Joshi2000}. To account for the fact that dense star clusters are subject to the tidal field of their host galaxy, we adopt an effective tidal mass-loss criterion that matches the tidal mass loss found in direct $N$-body simulations \citep{Chatterjee2010}.

Single and binary stars are respectively evolved with the \texttt{SSE} and \texttt{BSE} codes \citep{Hurley2000, Hurley2002,Chatterjee2010}, with up-to-date prescriptions for neutron star (NS) and black hole (BH) formation \citep{Fryer2001,Vink2001,Belczynski2002,Hobbs2005,Morscher2015,rodcomp2016}. In particular, two scenarios are considered for NS formation: iron core-collapse supernovae and electron-capture supernovae \citep{Ye2019}. In our simulations, the former receive natal kicks drawn from a Maxwellian with dispersion $\sigma=265\,\rm{km\,s}^{-1}$, the latter with dispersion $20\,\rm{km\,s}^{-1}$. Updated prescriptions for pulsar formation and evolution are also implemented \citep[see][for details]{Ye2019}. BHs are assumed to be formed with mass fallback and receive natal kicks by sampling from the same distribution as core-collapse supernovae NSs, but with kicks reduced in magnitude according to the fractional mass of fallback material \citep{Fryer2012,Morscher2015}. We also include prescriptions to account for pulsational-pair instabilities and pair-instability supernovae \citep{Belczynski2016b}.
    
Binary--single and binary--binary strong encounters are integrated using \texttt{Fewbody} \citep{Fregeau2004,Fregeau2007}, which includes gravitational radiation reaction for all encounters involving BHs \citep{Rodriguez2018a,Rodriguez2018b}. Collisions between stars during close encounters are treated in the sticky-sphere approximation, i.e. any pair of stars that pass close to one another are assumed to physically collide whenever their closest approach is smaller than the sum of their radii. Finally, we also take into account binary assembly through three-body-binary formation for every object \citep{Aarseth1976,HeggieHut2003,Morscher2015} and GW capture for interactions involving BHs \citep{Samsing2019_singlesingle}.

\subsection{Cluster models}
\label{subsect:clusmodels}

We use a set of $148$ independent cluster simulations. We consider different total number of particles (single stars plus binaries; $N=2\times10^5$, $4\times10^5$, $8\times10^5$, $1.6\times10^6$, and $3.2\times10^6$), initial cluster virial radius ($r_v/\rm{pc}=0.5,\,1,\,2,\,4$), metallicity ($Z/{\rm Z}_\odot=0.01,\,0.1,\,1$), and galactocentric distance\footnote{Assuming a Milky Way-like galactic potential \citep[e.g.,][]{Dehnen1998}} ($R_{\rm{gc}}/\rm{kpc}=2,\,8,\,20$).

We assume that all the models are initially described by a King profile, with initial King concentration parameter $W_0 = 5$ \citep{King1962}. Stellar masses are drawn from a canonical \citet{Kroupa2001} initial mass function (IMF) in the range $0.08-150\msun$. The primordial stellar binary fraction is fixed to $f_b=5\%$, with secondary masses drawn from a uniform distribution in mass ratio \citep[e.g.,][]{DuquennoyMayor1991}. Binary orbital periods are sampled from a log-uniform distribution \citep[e.g.,][]{DuquennoyMayor1991}, with the orbital separations ranging from near contact to the hard/soft boundary, while  binary eccentricities are drawn from a thermal distribution \citep[e.g.,][]{Heggie1975}.

Each simulation is evolved to a final time $T_{\rm H}=14$ Gyr, unless the cluster disrupts or undergoes a collisional runaway process \citep{kremer2020}. 

Primordial triples are not included in our cluster simulations. However, during strong binary--binary encounters, stable hierarchical triple systems can be formed \citep{rasio1995}. Limitations in \texttt{CMC} currently require these triples to be broken artificially at the end of the integration timestep. Nevertheless, whenever a stable triple is formed, its properties are logged, including the masses, stellar types, radii, and the semi-major axes and eccentricities for the inner and outer orbits\footnote{Note that, since these triple systems are \textit{de facto} destroyed in the Monte Carlo simulations, it is possible for the components of these triples to subsequently form new triple systems, when in reality they could survive for a significant period of time.}. Since we lack information regarding the mutual orientation of the two orbits, we sample their argument of periapsis $\omega_0$, cosine of the relative inclination $\cos I_0$, and orbital phases from a uniform distribution \citep{antoninietal2016}. To average out over these uncertainties, we realize this procedure $10$ times for each triple formed in each of the $148$ clusters presented in \citet{kremer2020}.

\section{Dissecting the origin of triples}
\label{sect:triporigin}

In this Section, we discuss the relevant formation channels of triples in star clusters, the characteristics of their progenitors, the formation times, and the recoil kicks that triple systems are imparted at the moment of formation. We label the inner and outer semi-major axes of the formed triples $a_{\rm in}$ and $a_{\rm out}$, respectively, the inner and outer eccentricities $e_{\rm in}$ and $e_{\rm out}$, respectively, the mass of the components of the inner binary $m_1$ and $m_2$ ($m_2<m_1$), the total mass of the inner binary $m_{\rm in}=m_1+m_2$, and the mass of the outer component $m_3$. The total mass of the triple is $m_{\rm t}=m_{\rm in}+m_3$, while the initial relative inclination of the inner and outer orbit is $i_0$. We label the remaining object $m_{\rm s}$ (fourth object leaving the system after the binary--binary interaction).

\subsection{Progenitors}
\label{subsect:progenitors}

\begin{figure} 
\centering
\includegraphics[scale=0.575]{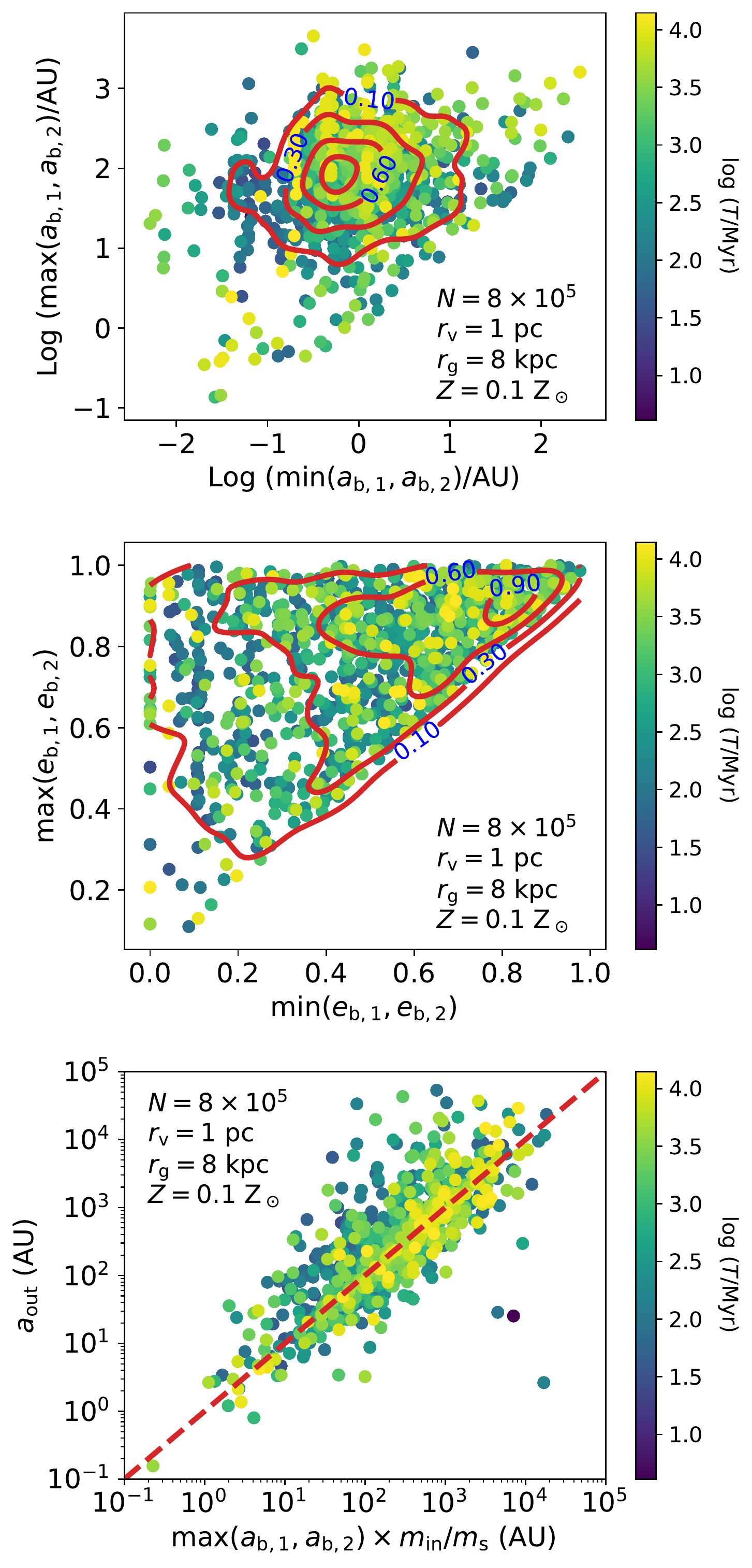}
\caption{Properties of binaries that lead to the formation of triple systems in binary--binary encounters for a cluster with initial number of stars $N=8\times 10^5$ ($r_{\rm v}=1$\,pc, $r_{\rm g}=8$\,kpc, $Z=0.1\ {\rm Z}_\odot$). Semi-major axes (top panel), eccentricities (middle panel), and outer semi-major axis as a function of the orbital elements of the binaries in the encounter (bottom panel) are shown. In the top two panels, the solid red lines represent the density contours of $10\%$, $30\%$, $60\%$, $90\%$ probability regions. The dashed red line in the bottom panel represents the $x=y$ line. The color map represents log formation time.}
\label{fig:prog}
\end{figure}

We find from our simulations that the majority of triple systems ($\sim 98.2\%$ of the overall triple population) are formed as a result of binary--binary encounters. In general, the probability of binary--binary encounters is \citep{binneytremaine2008}
\begin{equation}
\Gamma_{\rm bin-bin}\sim n_{\rm bin}^2 \sigma v_{\rm disp}\,,
\end{equation}
where $n_{\rm bin}$ is the density of binaries, $\sigma$ is the cross-section, and $v_{\rm disp}$ is the velocity dispersion. Since $n_{\rm bin}$ is largest in the core, the typical binary--binary encounter occurs in the core of dense star clusters. Of all the binary--binary encounters, the ones that successfully create triples involve two binaries of quite disparate sizes. Here, the tighter binary ejects a member of the wider binary and inserts itself, thus creating a stable hierarchical triple. The replaced object receives a dynamical recoil kick and is ejected from the encountering system\footnote{In some cases, its recoil velocity would be high enough to eject it from the cluster (see Sect~\ref{subsect:ejec}).}, while the captured one becomes the tertiary in the newly formed triple system.

We illustrate in Fig~\ref{fig:prog} the properties of binaries that lead to the formation of triple systems in binary--binary encounters for a cluster with initial number of stars $N=8\times 10^5$. The other initial cluster parameters are $r_{\rm v}=1$\,pc, $r_{\rm g}=8$\,kpc, $Z=0.1\ \mathrm{Z}_\odot$. In the top panel, we show the maximum of the semi-major axes ($a_{\rm b,1}$ and $a_{\rm b,2}$) of the two binaries that undergo the binary--binary encounter as a function of the minimum of them. We also overplot the probability density contours. We find that the bulk of the interactions that produce a triple include two binaries, of which one is wider than the other by $\sim 2$ orders of magnitude. This confirms our picture, where triples typically form when a binary replaces one of the components of a wider binary. We also show in Fig~\ref{fig:prog} (middle panel) the maximum eccentricities of the two binaries undergoing the binary--binary encounters ($e_{\rm b,1}$ and $e_{\rm b,2}$) as a function of the minimum of them. Since encounters thermalize the distribution of the eccentricities of the progenitors \citep{Heggie1975}, most of the binaries that produce triples are highly eccentric.

Since the typical triple-producing binary--binary encounter involves a tight binary exchanging into a wide binary, we expect the outer semi-major axis distribution of the outer semi-major axis of the triples to be related to the orbital elements of the ionized binaries. From energy conservation,
\begin{equation}
\frac{m_{\rm in} m_3}{a_{\rm out}}\sim \frac{m_3 m_{\rm s}}{\max(a_{\rm b_1},a_{\rm b_2})}\,,
\end{equation}
where $m_{\rm s}$ is the mass of the replaced component in the wider binary \citep{sigphi1993}. Therefore, the outer semi-major axis of the triple is linearly related to the semi-major axis of the wider binary through
\begin{equation}
a_{\rm out}\sim \frac{m_{\rm in}} {m_{\rm s}}\max(a_{\rm b_1},a_{\rm b_2})\,.
\end{equation}
We show this in the bottom panel of Fig~\ref{fig:prog}. As expected, the majority of the systems lie on the $x=y$ line. Triples that are outliers with respect to this simple scaling are systems formed during resonant encounters, where the energy is redistributed in a more complex way during multiple passages and interactions among the four objects (two binaries) involved in the encounter.

\subsection{Cluster and triple properties}
\label{subsec:clustrip}

\begin{figure*} 
\centering
\includegraphics[scale=0.6]{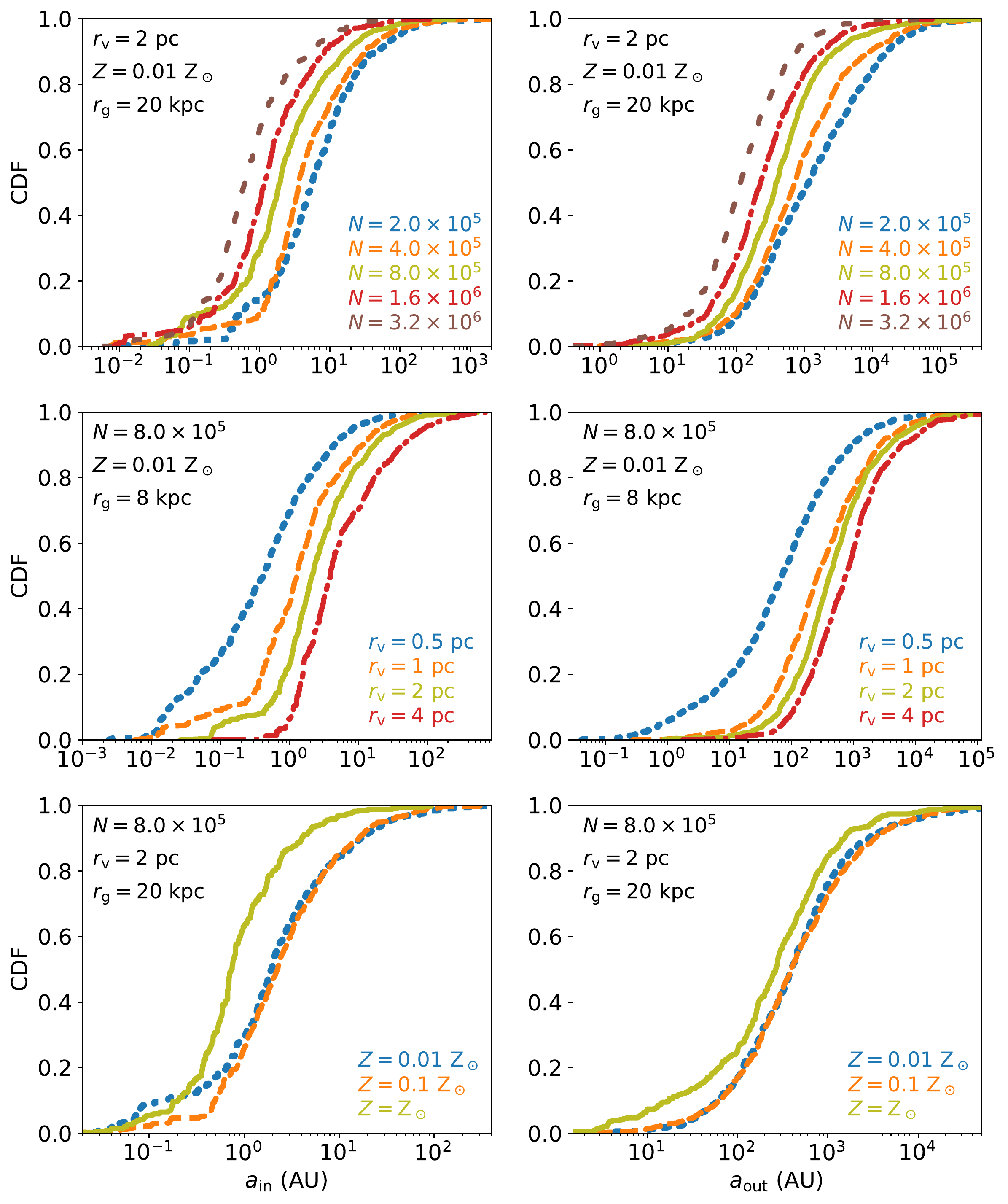}
\caption{Cumulative distribution functions of inner ($a_{\rm in}$; left panels) and outer ($a_{\rm out}$; right panels) semi-major axis of triples in clusters of various initial numbers of stars (top panels), virial radii (central panels), and metallicities (bottom panels).}
\label{fig:ainaout}
\end{figure*}

The initial conditions of the parent cluster set the distribution of the orbital elements of the formed triple systems. We show this in Figure~\ref{fig:ainaout}, where we plot the cumulative distribution functions of inner and outer semi-major axes of triples in clusters of various initial numbers of stars, virial radii, and metallicities.

In the top panel of Figure~\ref{fig:ainaout}, we illustrate the cumulative distribution function of triples in clusters of different initial number of stars ($N=2\times 10^5$--$3.2\times 10^6$) and $r_{\rm v}=2$\,pc, $Z=0.01\,\zsun$, $r_{\rm g}=20$\,kpc. Triples that form in larger clusters tend to have smaller inner and outer semi-major axes. We find that $\sim 50\%$ of the systems have $a_{\rm in}/\rm{au}\lesssim(0.6,1,2,4,5)$ and $a_{\rm out}/\rm{au}\lesssim(1,2,4,7,10)\times10^2$ for $N=(32,16,8,4,2)\times10^5$, respectively. This comes from the fact that binaries that undergo binary--binary scattering and produce a triple system are tighter in more massive clusters. In these environments, stellar densities are typically higher than in less massive clusters and wide binaries are ionized by encounters with stars and compact objects. 

We plot in the middle panel of Figure~\ref{fig:ainaout} the cumulative distribution function of triples in clusters of different initial virial radii $r_{\rm v}/\rm{pc}\in [0.5,4]$ and $N=8\times 10^5$, $r_{\rm g}=8$\,kpc, $Z=0.01\,\zsun$. Triples that form in clusters with larger virial radii tend to have wider inner and outer orbits. We find that $\sim 50\%$ of the triple systems have $a_{\rm in}/\rm{au}\lesssim (0.3,1,2,4)$ and $a_{\rm out}/\rm{au}\lesssim (70,250,400,700)$ for $r_{\rm v}/\rm{pc}=(0.5,1,2,4)$, respectively. This is expected since clusters with smaller values of $r_{\rm v}$ typically have a higher density and velocity dispersion. Thus, the progenitor binaries (that later undergo binary--binary encounter to form triples) have to be more compact in order to remain bound after encounters with stellar or compact objects. 

Finally, in the bottom panel of Figure~\ref{fig:ainaout}, we plot the cumulative distribution function of triples in clusters of different initial metallicities $Z/\zsun=(0.01,0.1,1),$ with $N=8\times 10^5$, $r_{\rm v}=2$\,pc, and $r_{\rm g}=20$\,kpc. Triples that form in higher metallicity clusters tend to have smaller inner and outer semi-major axes. We find that $\sim 50\%$ of the triple systems have $a_{\rm in}/\rm{au}\lesssim (2,2,0.7)$ and $a_{\rm out}/\rm{au}\lesssim (400,400,250)$ for $Z/\zsun=(0.01,0.1,1)$, respectively. This can be related to the BH-burning process \citep{Kremer2019d}. BHs in metal-rich clusters are low-mass and do not inject as much energy into the BH-burning process as BHs in metal-poor clusters. Thus, these clusters typically have higher densities and dispersion velocities. As a result, metal-poor clusters allow wider binaries to form triples compared to metal-rich clusters.

\subsection{Cluster properties and formation times}
\label{subsect:tform}

Triple systems are not formed uniformly in time. Rather, they track the evolutionary paths of the parent cluster. The clock of a star cluster is essentially set by its half-mass relaxation time \citep{Spitzer1987}
\begin{equation}
t_{\rm rh}\sim \frac{N^{1/2} r_{\rm v}^{3/2}}{\langle m\rangle^{1/2} G^{1/2} \ln \Lambda}\,,
\label{eqn:trh}
\end{equation}
where $\langle m\rangle$ is the average mass in the cluster and $\ln\Lambda$ is the Coulomb logarithm. As discussed in greater detail in \citet{Kremer2019a}, the initial cluster size, set by its initial virial radius, is the key parameter which determines the ultimate fate of a cluster and its BH population (``BH-burning`` mechanism). Clusters with smaller initial $r_{\rm v}$ have shorter relaxation times and have a dynamical clock that runs faster compared to clusters born with larger initial virial radius. These clusters could eject the majority of their BH population over their lifetime and appear as core-collapsed clusters.

\begin{figure} 
\centering
\includegraphics[scale=0.625]{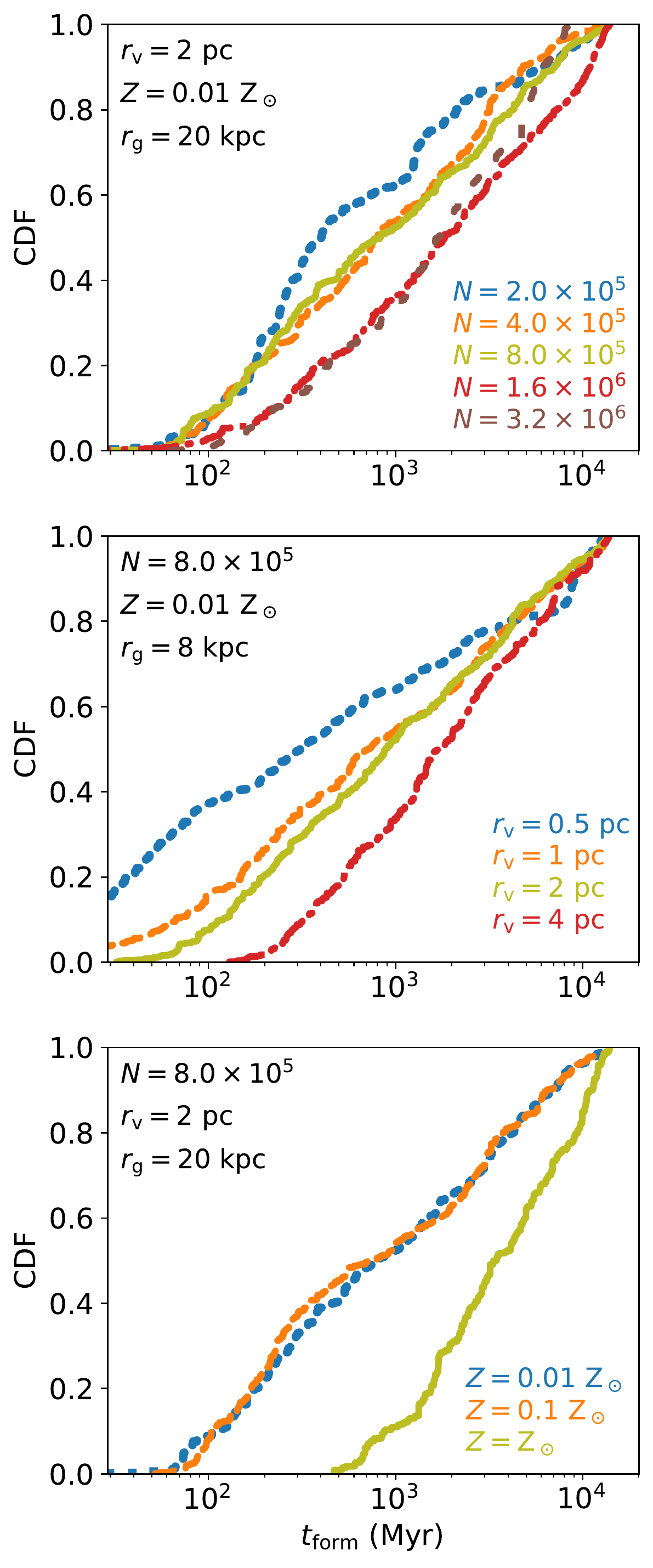}
\caption{Formation time ($t_{\rm form}$) of triples in clusters of various initial numbers of stars (top panel), virial radii (central panel), and metallicities (bottom panel).}
\label{fig:tform}
\end{figure}

In Figure~\ref{fig:tform}, we plot the formation time ($t_{\rm form}$) of triples in clusters of various initial numbers of stars, virial radii, and metallicities (same as Figure~\ref{fig:ainaout}).

In the top panel, we show the cumulative distribution function of triples in clusters of different initial numbers of stars $N\in [2\times10^5,3.2\times10^6]$, $r_{\rm v}=2$\,pc, $Z=0.01\,\zsun$, and $r_{\rm g}=20$\,kpc. As expected from Eq.~\ref{eqn:trh}, larger star clusters have longer evolutionary timescales. Hence, triples are assembled through binary--binary scatterings later compared to smaller clusters. We find that $\sim 50\%$ of the triples are assembled at $t_{\rm form}\lesssim 0.2$ Gyr ($\sim 0.25\,t_{\rm rh}$) for $N=2\times 10^5$, while $\sim 50\%$ of the triples are assembled at $t_{\rm form}\lesssim 2$ Gyr ($\sim 0.5\,t_{\rm rh}$) for $N=3.2\times 10^6$.

We plot in the middle panel of Figure~\ref{fig:tform} the cumulative distribution function of triples in clusters of different initial virial radii $r_{\rm v}/\rm{pc}\in[0.5,4]$, $N=8\times 10^5$, $Z=0.01\,\zsun$, $r_{\rm g}=8$\,kpc. As discussed, the initial cluster size sets the dynamical clock of a stellar cluster. Among the four represented clusters, the ones with $r_{\rm v}=0.5$\,pc and $r_{\rm v}=1$\,pc are core-collapsed \citep[see Figure~5 in][]{kremer2020}. Clusters with small initial virial radii form most of the triple systems much more quickly than clusters with larger initial sizes.

Finally, in the bottom panel of Figure~\ref{fig:tform}, we show the cumulative distribution function of triples in clusters of different initial metallicities $Z/\zsun\in[0.01,1]$, $N=8\times 10^5$, $r_{\rm v}=2$\,pc, and $r_{\rm g}=20$\,kpc. Star clusters with smaller metallicities form more massive BHs than clusters with higher metallicities \citep[see Figure~1 in][]{kremer2020}. Therefore, BHs are dynamically processed faster as the dynamical clock of the host cluster runs faster in the former case, producing triples on shorter timescales.

\subsection{Recoils and ejections}
\label{subsect:ejec}

Binary--binary exchange encounters impart recoil kicks to any triples they produce. \citet{leigh2016} showed that the ejection velocity of the single escaper ($m_{\rm s}$) in such an encounter is well described by the distribution\footnote{This assumes that the initial angular momentum is negligible. For a general discussion, see \citet{valt2006}.}
\begin{equation}
f(v_{\rm ej,s})dv_{\rm ej,s}=\frac{3|E|^{2}\mathcal{M}v_{\rm ej,s}}{\left(|E|+\frac{1}{2}\mathcal{M}v_{\rm ej,s}^{2}\right)^3}dv_{\rm ej,s}\,,
\end{equation}
where
\begin{equation}
\mathcal{M}=\frac{m_s (m_s+m_{\rm in})}{m_{\rm in}}\ 
\end{equation}
and $|E|$ is the total initial energy. From the conservation of linear angular momentum, the recoil velocity of the triple is
\begin{equation}
v_{\rm rec}=\frac{m_s}{m_t} v_{\rm ej,s}\,.
\end{equation}
This recoil kick can be large enough to eject the triple from the core (where it will eventually sink back as a result of dynamical friction) or even from the cluster.

\begin{figure*} 
\centering
\includegraphics[scale=0.65]{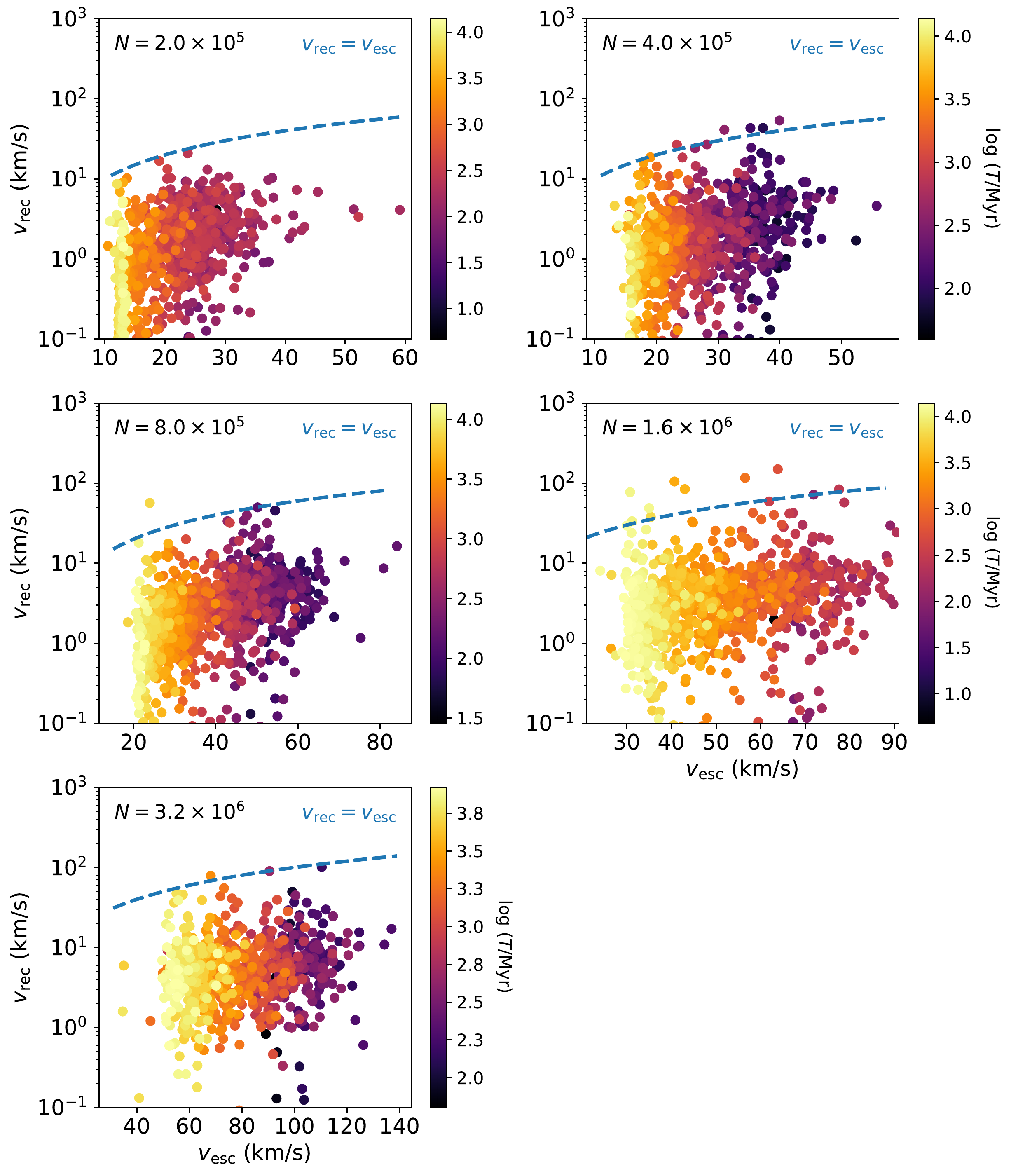}
\caption{Recoil velocity $v_{\rm rec}$ of the triple systems assembled in the cores of star clusters of various initial numbers of stars $N$ ($r_{\rm v}=2$\,pc, $r_{\rm g}=20$\,kpc, $Z=0.01\,{\rm Z}_\odot$), as a function of the cluster escape speed $v_{\rm esc}$ at the moment of formation. The dot-dashed blue line represents $v_{\rm rec}=v_{\rm esc}$. Color code: log formation time.}
\label{fig:vejvesc}
\end{figure*}

We use the data recorded on binary--binary scatterings that lead to the formation of stable triple systems during the cluster's lifetime to compute $v_{\rm rec}$. In Figure~\ref{fig:vejvesc}, we show the recoil velocity $v_{\rm rec}$ of the triple systems assembled in the cores of star clusters of various initial $N$ (for fixed $r_{\rm v}=2$\,pc, $r_{\rm g}=20$\,kpc, and $Z=0.01\,{\rm Z}_\odot$) as a function of the cluster escape speed $v_{\rm esc}$ at the moment of formation. For these clusters, we find that $\lesssim 1\%$ of the formed triples could escape the clusters due to dynamical recoil kicks (if they do not encounter other stars or compact objects). Moreover, the escaping systems tend to be ejected from the cluster at later times, when the cluster escape speed decreases to lower values. Most of the triples will not leave the cluster. Rather, they will be kicked on elongated orbits out of the cluster core. As they are more massive than the average star, they would sink back to the cluster core on a dynamical friction timescale
\begin{equation}
t_{\rm df}\sim \frac{\langle m\rangle}{m_{\rm t}}t_{\rm rh}\,,
\label{eqn:tdf}
\end{equation}
where $m_{\rm t}=m_{\rm in}+m_3$ is the total mass of the triple.

\section{Demographics}
\label{sect:demogr}

In this section, we discuss how the parent cluster initial conditions shape the orbital properties of the formed triples and describe their demographics.

We are interested in triples that are hierarchically stable. While simulating strong encounters inside CMC, triple stability is checked using the stability criteria given by \citet{mard01},
\begin{equation}
\frac{a_{\rm out}}{a_{\rm in}} \mathcal{R}\left(e_{\rm out},\frac{m_{\rm out}}{m_{\rm in}}\right)\geq 2.8\,,
\label{eqn:stabts}
\end{equation}
where
\begin{eqnarray}
\mathcal{R}\left(e_{\rm out},\frac{m_{\rm out}}{m_{\rm in}}\right)&=& \left[\left(1+\frac{m_{\rm out}}{m_{\rm in}}\right)\frac{1+e_{\rm out}}{\sqrt{1-e_{\rm out}}} \right]^{-2/5}\nonumber\\
& \times & (1-e_{\rm out}) \left(1-\frac{0.3\, i_0}{180^\circ}\right)\,.
\label{eqn:stabts2}
\end{eqnarray}

We subdivide the triple population into four categories, such that the stellar types of the two components $k_1$ and $k_2$ of the inner binary \citep[see][]{Hurley2000} are always $k_1\le k_2$:
\begin{itemize}
    \item triples with a main-sequence (MS) star in the inner binary;
    \item triples with a giant (G) star in the inner binary;
    \item triples with a WD in the inner binary;
    \item triples with a NS or BH in the inner binary.
\end{itemize}
Among the systems with an inner BH-BH binary, we also consider triples where all the components are BHs, that we label BH-BH-BH.

As a general trend, we find that a cluster typically assembles hundreds of triples with an inner BH-BH binary (of which $\sim 70\%$--$90\%$ have a BH as tertiary) or an inner MS-BH binary. Additionally, tens of triples with an inner MS-MS or WD-BH are produced. However, only clusters with $r_{\rm v}\le 1$\,pc efficiently assemble triples with an inner binary comprised of a MS-WD or WD-WD, and produce $\sim 10$ times more systems with an inner MS-MS binary. Again, this is a natural consequence of the BH burning process \citep{Kremer2019d}, since only clusters with small initial virial radii are able to eject most of their BH population, thus allowing lighter objects to sink to their innermost regions and efficiently produce triples. Moreover, we find that $\sim 50\%$ of the overall triple population from our simulations consists of systems where all the components are BHs. Roughly 10\% of the systems take the form of a binary BH with a non-BH third companion and $\sim38\%$ the form of an inner binary with at least one MS star. Other triples constitute the remaining $\sim 2\%$. Tables~\ref{table:models}-\ref{table:models2} summarize all the different triples formed in each cluster simulation in our ensemble, subdivided into the above described categories.

\subsection{Gravitational wave captures and mergers during triple formation}
\label{subsec:disrpgwcap}

\begin{figure} 
\centering
\includegraphics[scale=0.55]{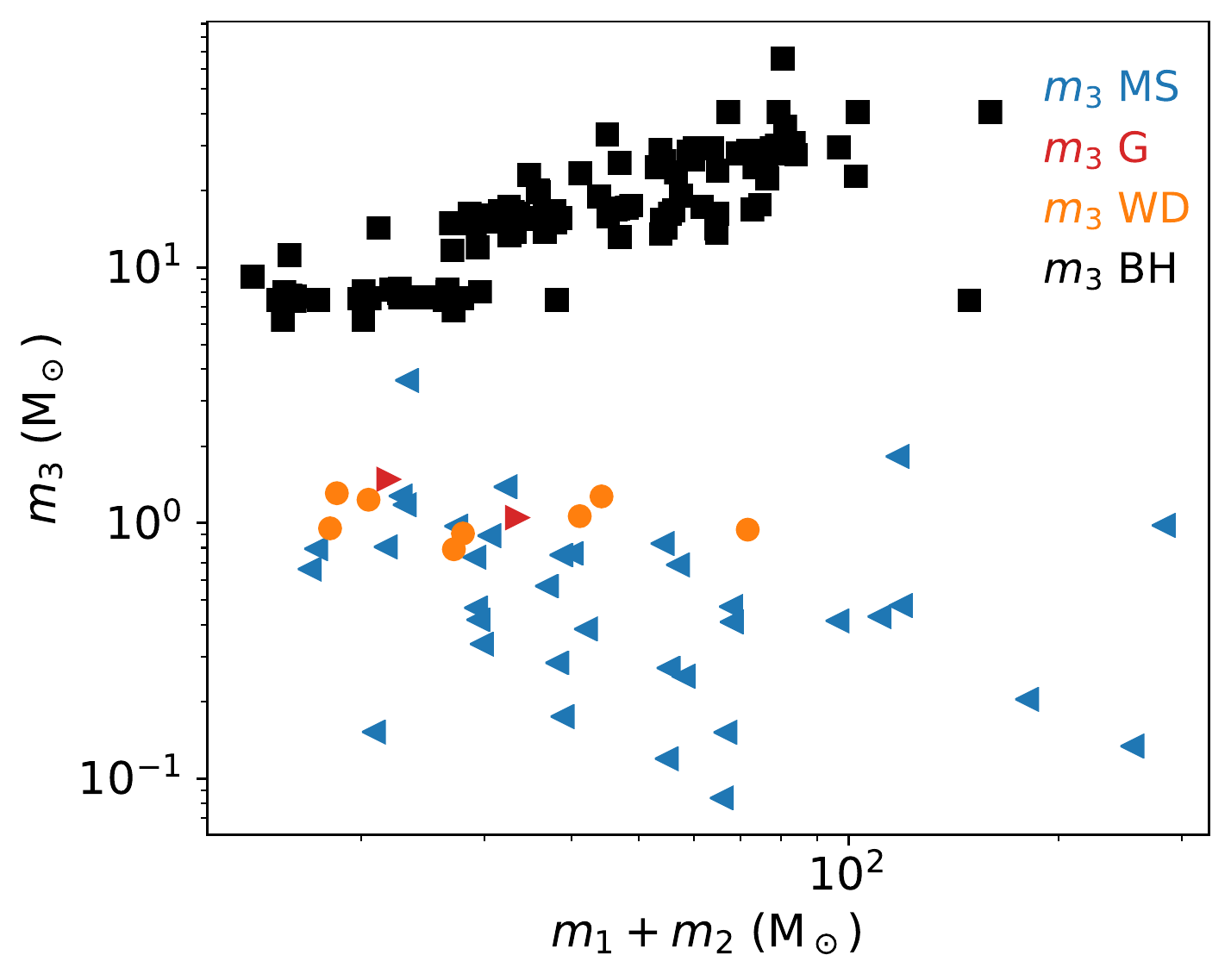}
\caption{Outer mass as a function of the total mass of the inner binary of the triple systems that form through GW captures during binary--single encounters. The binary, which becomes the inner binary of the triple, is always a binary BH.}
\label{fig:proggw}
\end{figure}

\begin{figure} 
\centering
\includegraphics[scale=0.55]{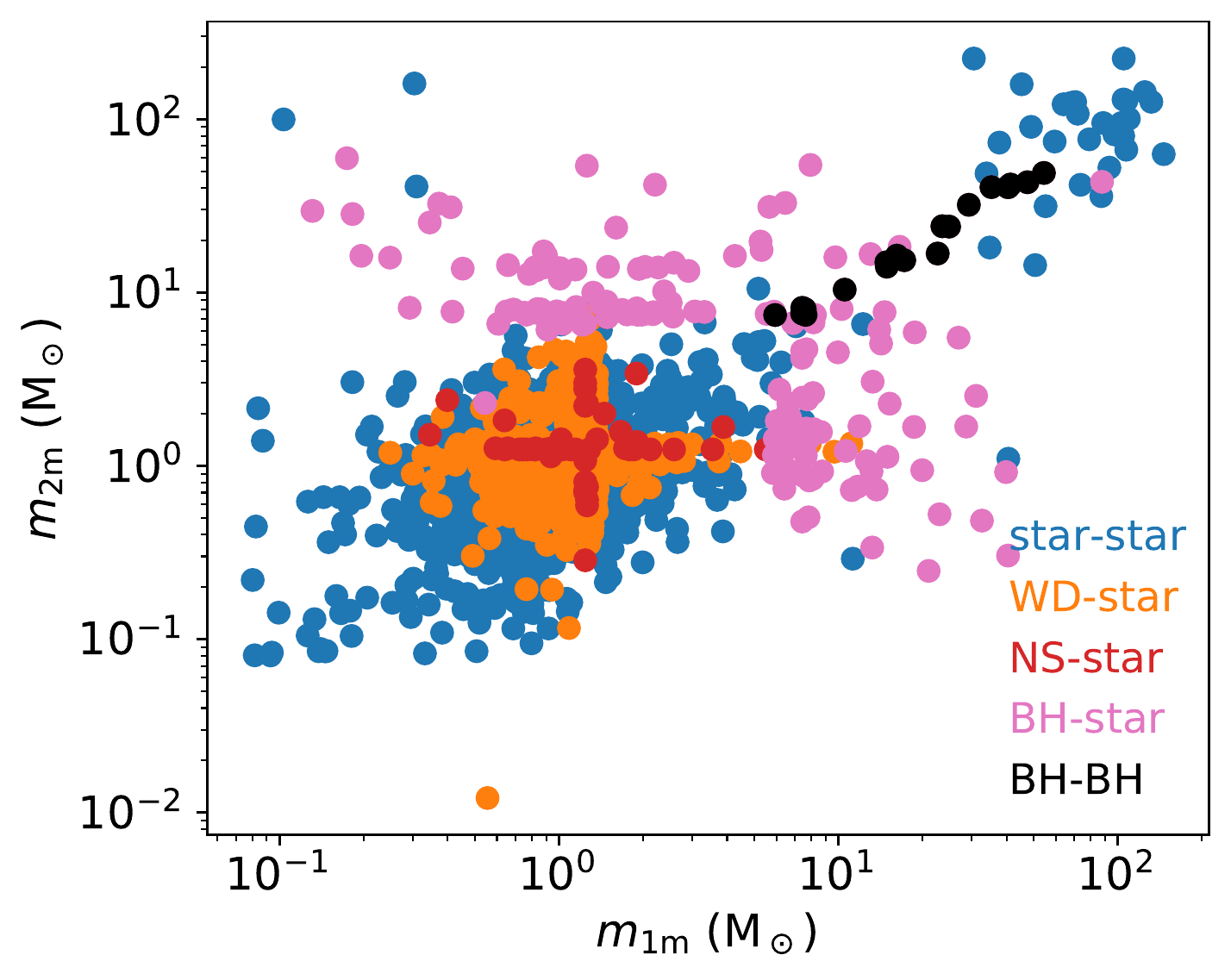}
\caption{Masses of the components ($m_{\rm 1m}$ and $m_{\rm 2m}$) that merge during binary--binary encounters that lead to triple formation. Different colors represent different stellar and compact object types.}
\label{fig:progmerg}
\end{figure}

A handful of triple systems ($\sim 0.1\%$ of the overall population) are formed during binary--single encounters as a result of GW captures \citep{Samsing2019_singlesingle}. In this process, the single has the chance to pass sufficiently close to the binary to dissipate some energy via GW radiation, thus remaining bound to the binary itself. For all triples assembled this way in our simulations, we show in Figure~\ref{fig:proggw} the outer mass as a function of the inner binary's total mass. We find that the binary that intervenes in the process, which later becomes the inner binary of the triple, is always a binary BH. The majority of the triples formed through GW captures are made up of three BHs, while a few systems have a star (either MS or G) or WD as the outer companion. We find no GW capture systems with a NS outer companion.

During the binary--binary encounters that produce a triple system, two of the objects can pass close enough to merge. This can occur in multiple ways: collision and merger of two stars (MS or G), tidal disruption of stars by a compact object, and merger of two compact objects. In Figure~\ref{fig:progmerg}, we plot for all simulations the masses of the components ($m_{\rm 1m}$ and $m_{\rm 2m}$) that merge during binary--binary encounters which yield triples ($\sim 1.7\%$ of their overall population). Different colors represent different stellar and compact object types. Among the stars that collide, we find that $\sim 90\%$ and $\sim 10\%$ of the mergers are with MS or G stars, respectively.In the standard scenario for triple formation, the tighter binary ejects the single star it replaces, but no ejection occurs in this process.

\subsection{Stability and softness}
\label{subsec:softstab}

\begin{figure*} 
\centering
\includegraphics[scale=0.6]{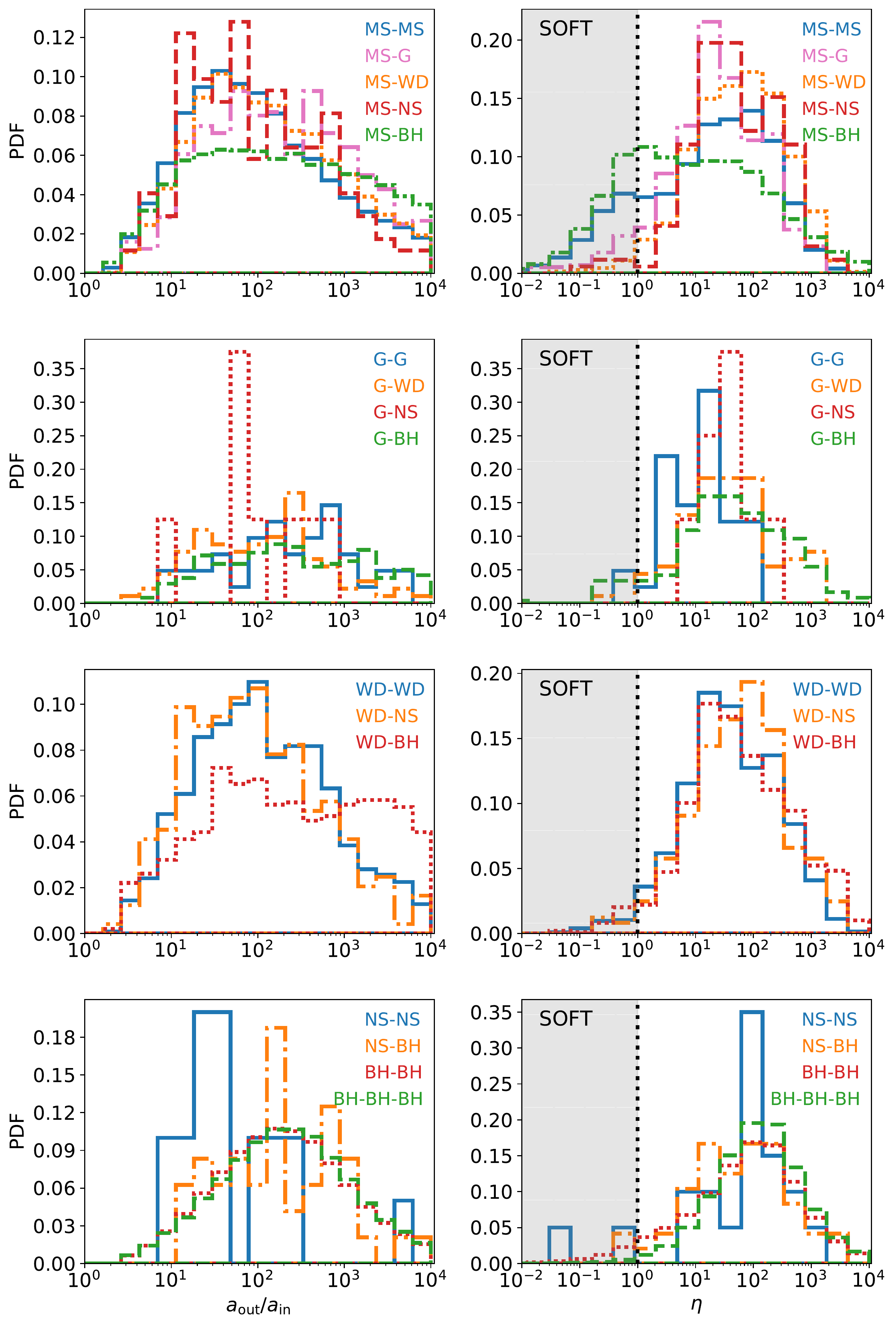}
\caption{Probability distribution function of the ratio of outer and inner semi-major axes ($a_{\rm out}/a_{\rm in}$; left) and the softness parameter ($\eta$; right) of all triple populations formed in our $148$ cluster simulations. Top panels: triples with a MS star plus a companion in the inner binary. Central-top panels: triples with a G star plus a companion in the inner binary. Central-bottom panels: triples with a WD plus a companion in the inner binary. Bottom panels: triples with a NS or BH plus a companion in the inner binary.}
\label{fig:softstable}
\end{figure*}

\begin{figure*} 
\centering
\includegraphics[scale=0.6]{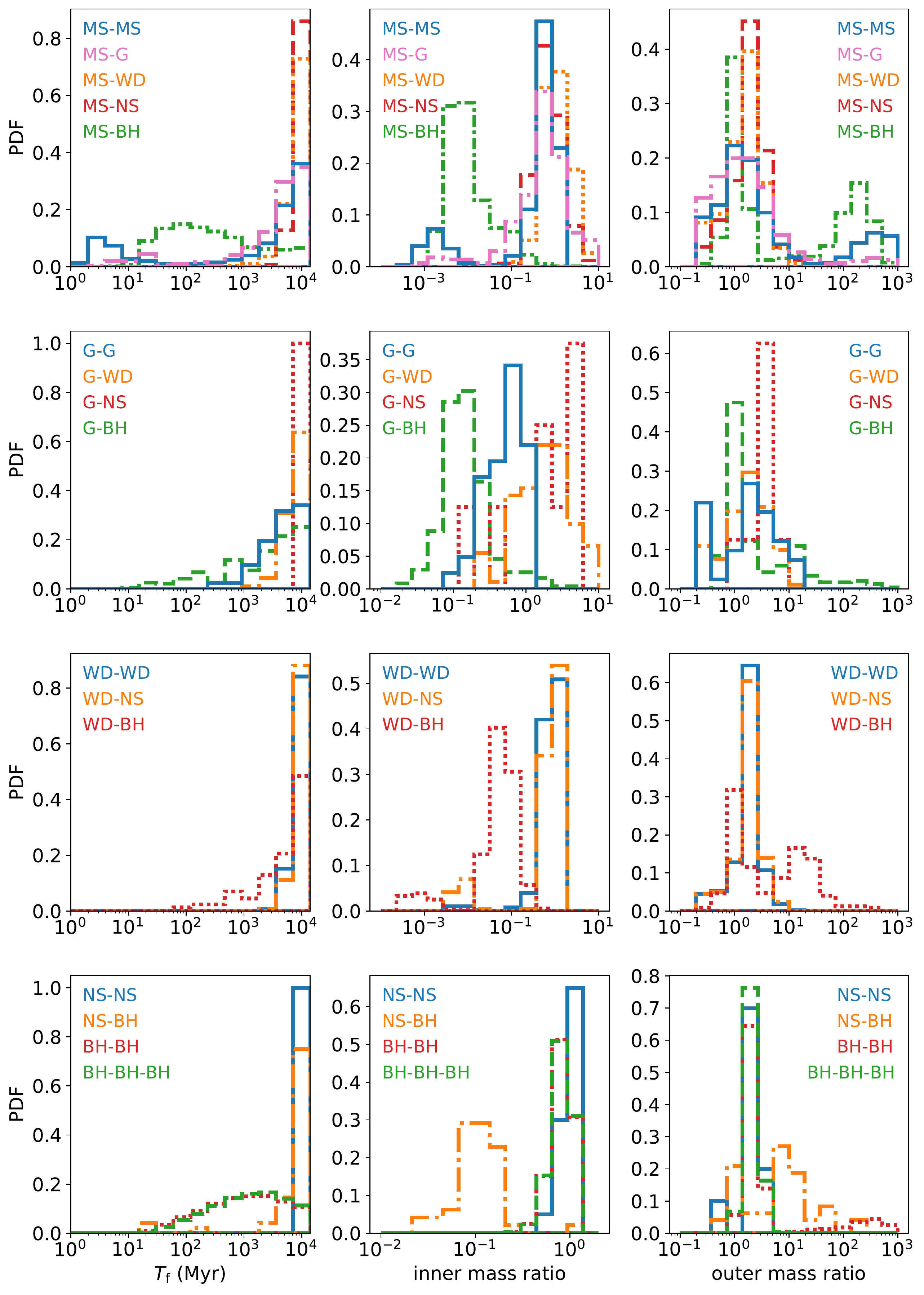}
\caption{Probability distribution function of the formation time ($T_{\rm f}$; left), the inner mass ratio (center), and the outer mass ratio (right) of all the triple populations formed in our $148$ cluster simulations. Top panels: triples with a MS star plus a companion in the inner binary. Central-top panels: triples with a G star plus a companion in the inner binary. Central-bottom panels: triples with a WD plus a companion in the inner binary. Bottom panels: triples with a NS or BH plus a companion in the inner binary.}
\label{fig:tinmoutm}
\end{figure*}

We define the softness parameter \citep{Heggie1975}
\begin{equation}
\eta\equiv\frac{Gm_{\rm in}m_3}{2 a_{\rm out}\langle m\rangle v_{\rm disp}^2}\,,    
\end{equation}
where $\langle m \rangle$ and $v_{\rm disp}$ are the average mass in the cluster and the cluster velocity dispersion, respectively. Triples that have $\eta\ll 1$ are referred to as `soft' and will become even softer on average, until they are disrupted by the background population. Triples with $\eta\gg 1$ are referred to as `hard' and tend to become even harder by interacting with cluster stars \citep{Heggie1975}.

We illustrate in the left panels of Figure~\ref{fig:softstable} the probability distribution function of the ratio of the outer and inner semi-major axes of all triple populations formed in our $148$ cluster simulations. We find that the majority of the systems have $a_{\rm out}/a_{\rm in}\gtrsim 10$, regardless of the composition of the inner binary. We also show in Figure~\ref{fig:softstable} the probability distribution function of the softness parameter $\eta$ (right panels) of all triples formed in the simulations. We find quite generally that triple populations have $\eta\gg1$, with only a small tail of soft triples and a main peak at $\eta \sim 100$.

\subsection{Formation time, inner mass ratio, outer mass ratio}
\label{subsec:softstab}

We show in Figure~\ref{fig:tinmoutm} the probability distribution function of the formation time (left), the inner mass ratio (center), and the outer mass ratio (right) of all the triple populations formed in the simulations.

As a common trend, we find that triples whose inner binary has at least one BH typically form on a shorter timescale compared to other triples. This can be understood in terms of the BH-burning mechanism \citep{Kremer2019d}. In this process, strong dynamical encounters between the BHs act as an energy source for the rest of the cluster. Thus, BHs tend to occupy the innermost and densest parts of the cluster, where most of the binary--binary interactions take place, preventing other components from efficiently segregating there. As a result, triples whose inner binary does not contain a BH tend to form on longer timescales, when most of the BHs have been processed and have left the cluster.

For MS stars, we define the inner mass ratio $m_{\rm MS}/m_{\rm comp}$ as the ratio between the MS star's mass ($m_{\rm MS}$) and that of its companion ($m_{\rm comp}$). If there are two MS stars in the inner binary, we define the inner mass ratio as $m_{\rm MS,2}/m_{\rm MS,1}$, with $m_{\rm MS,1}>m_{\rm MS,2}$. The same applies to G stars, WDs, NSs, and BHs. The outer mass ratio is simply defined as the ratio between the total mass of the inner binary and the mass of the tertiary. Interestingly, we find that the inner mass ratio is usually peaked at $\sim 1$, unless the system only has one BH in the inner binary. The distribution of outer mass ratios is also nearly peaked at $\sim 1$, except for systems with an inner binary comprised of a MS-MS, MS-BH, WD-BH, or NS-BH. The secondary peaks at $\sim 10$--$100$ corresponds to a low-mass stellar tertiary.

\subsection{Survivability}
\label{subsect:surviv}

\begin{figure} 
\centering
\includegraphics[scale=0.475]{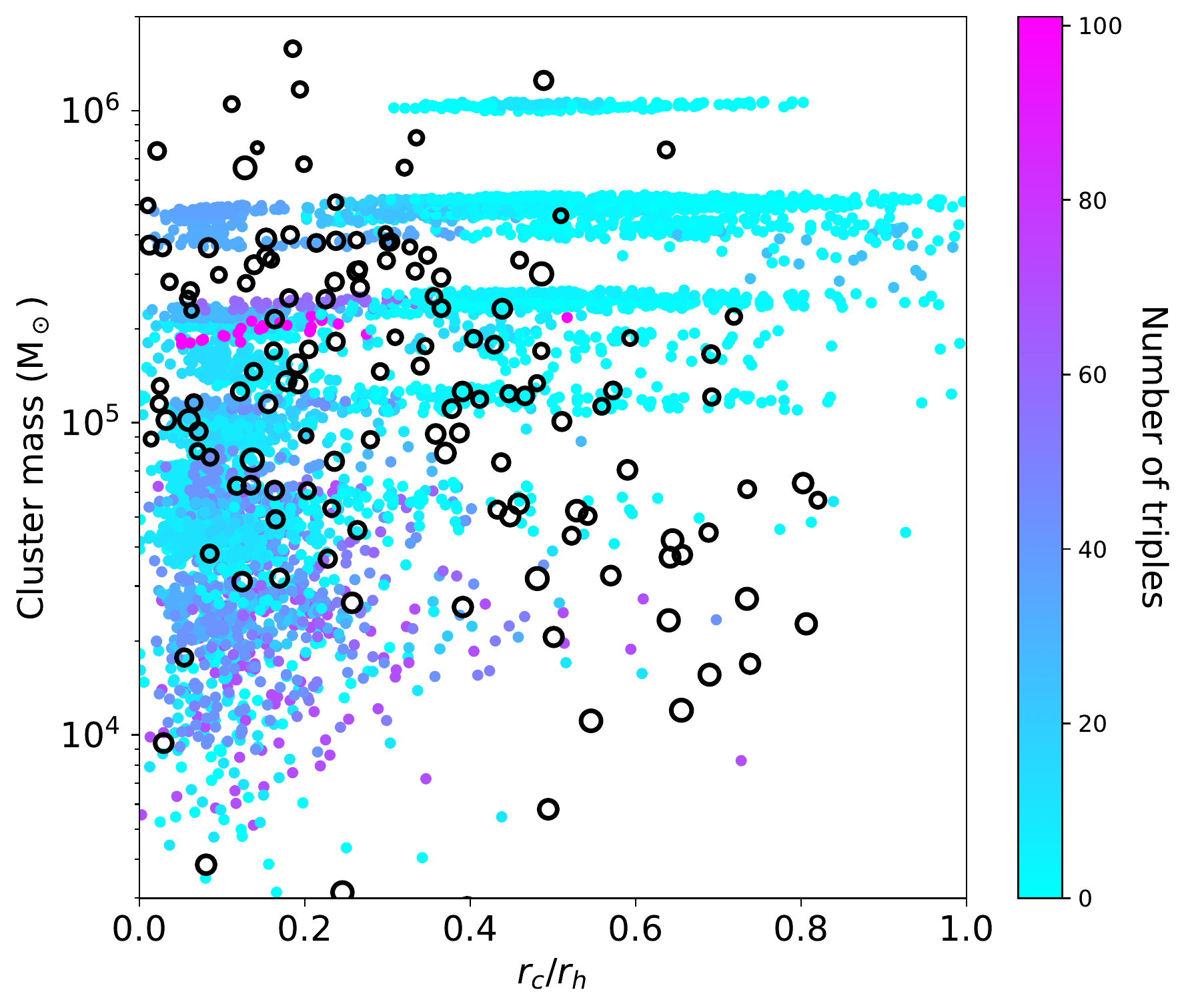}
\caption{All late-time snapshots ($10$--$13$ Gyr) for model clusters masses and concentrations (colored points). Milky Way clusters (black points) are taken from \citet{Baumgardt2018}, with the size of each black point corresponding to the integrated V-band magnitude of each cluster \citep[][larger symbols denote clusters that are best observed]{Harris1996}. Color code: number of triples with at least one luminous component that survive at present day unperturbed in the cluster.}
\label{fig:ntriplemw}
\end{figure}

In the dense stellar environment of star clusters, triple systems may be perturbed through encounters with other passing objects. Such encounters will alter the orbital properties of the triple significantly or even disrupt it. This process happens on a typical timescale\footnote{Quantities $x_a$ are expressed with physical units $u$ as $x_{\mathrm{a},u} \equiv x_{\mathrm{a}}/u$, so that $x_{\mathrm{a},u}$ is dimensionless.} \citep{binneytremaine2008,Ivanova2008}
\begin{eqnarray}
T_{\rm enc}&=&8.5\times 10^{12}\,\mathrm{yr}\,P_{\rm out,d}^{-4/3} m_{\rm trip,\msun}^{-2/3} \sigma_{10\kms}^{-1} n_{10^5\,\mathrm{pc}^{-3}}^{-1}\times\nonumber\\
&\times&\left[1+913\frac{m_{\rm trip,\msun}+\langle m\rangle_{\msun}}{2P_{\rm out,d}^{2/3} m_{\rm trip,\msun}^{1/3} \sigma_{10\kms}^{2}}\right]^{-1}\ ,
\label{eqn:tenc}
\end{eqnarray}
where $P_{\rm out}$ is the orbital period of the outer orbit and $\langle m\rangle$ is the average stellar mass in the cluster. 

We show in Figure~\ref{fig:ntriplemw} all late-time snapshots ($10$--$13$ Gyr) for model clusters compared to Milky Way clusters. The latter are taken from \citet{Baumgardt2018} and represented such that their size is proportional to the integrated V-band magnitude of each cluster \citep[][]{Harris1996}. Thus, larger symbols denote clusters that are best observed. In color code, we represent the number of triples with at least one luminous (observable) component that survive in the cluster, i.e. triples whose encounter timescales are long enough to remain unperturbed. We find that clusters are on average expected to host tens of luminous triples at present.

\section{Triple-assisted mergers: transients and gravitational waves}
\label{sect:mergers}

In this section, we discuss the LK mechanism that takes place in triple systems. We then apply an analytical formalism to compute the maximum eccentricity attained by the triples formed in our simulations (subdivided as described in Section~\ref{sect:demogr}) and to infer the fraction of systems that result in a merger, a transient phenomenon, or GW emission by the LK mechanism.

\subsection{Lidov-Kozai mechanism}
\label{subsect:lkmech}

A triple system made up of an inner binary that is orbited by an outer companion undergoes LK oscillations in eccentricity whenever the initial mutual inclination of the inner and outer orbits is in the range $40^{\circ}\lesssim i_0\lesssim 140^{\circ}$ \citep[][quadrupole order of approximation]{lid62,koz62}. During these cycles, the eccentricity and inclination of the inner orbit can experience periodic oscillations on a secular quadrupole LK timescale
\begin{equation}
T_{\rm LK}=\frac{8}{15\pi}\frac{m_{\rm trip}}{M_{3}}\frac{P_{\rm out}^2}{P_{\rm in}}\left(1-e_{\rm out}^2\right)^{3/2}\,.
\label{eqn:tlk}
\end{equation}
In the previous equation, $P_{\rm in}$ and $P_{\rm out}$ are the orbital periods of the inner and outer binaries, respectively. We note that the exact size of the LK inclination window depends also on the physical parameters of the three objects, thus varying from case to case \citep[e.g.,][]{grish2018}. On this typical timescale, the relative inclination of the inner orbit and outer orbit slowly increases while the orbital eccentricity of the inner orbit decreases, and vice versa, conserving angular momentum \citep[see][for a review]{naoz2016}. The inner eccentricity can reach almost unity during LK cycles, which is typically achieved in the case $i_0\sim 90^\circ$. 

Whenever the outer orbit is eccentric (octupole order of approximation), the inner eccentricity can reach almost unity even if the initial inclination lies outside of the window $\sim 40^\circ$-$140^\circ$ \citep{naoz13a}. This happens over the octupole timescale
\begin{equation}
T_{\rm oct}=\frac{1}{\epsilon}T_{\rm LK}\,,
\label{eqn:tlkoct}
\end{equation}
where the octupole parameter is defined as 
\begin{equation}\label{oc1}
\epsilon={m_1-m_2\over m_1+m_2}\frac{a_{\rm in}}{a_{\rm out}}\frac{e_{\rm out}}{1-e_{\rm out}^2}\,.
\end{equation}

Nevertheless, LK cycles can be suppressed by additional sources of precession \citep[e.g.,][]{fabt2007,naoz13a}, such as non-dissipative tides, that operate on a timescale \citep{kisel1998,egg2001}
\begin{eqnarray}
T_{\rm tide}&=&\frac{8a_{\rm in}^{13/2}}{15(Gm_{\rm in})^{3/2}}\frac{(1-e_{\rm in}^2)^5}{8+12e_{\rm in}^2+e_{\rm in}^4}\nonumber\\
&\times& \left[2\frac{m_2}{m_1}k_1 R_1+2\frac{m_1}{m_2}k_2 R_2\right]^{-1}\,,
\label{eqn:ttidal}
\end{eqnarray}
where $k_1$, $R_1$ and $k_2$, $R_2$ are the apsidal motion constants and radii of the two stars in the binary \citep{hut1981}, respectively, or general relativistic (GR) precession, that operates on a typical timescale \citep{Peters1964}
\begin{equation}
T_{\rm GR}=\frac{a_{\rm in}^{5/2}c^2(1-e_{\rm in}^2)}{3G^{3/2}(m_{\rm 1}+m_{\rm 2})^{3/2}}\,.
\end{equation}

To compute the maximum eccentricity $e_{\max}$ attained by triples, we use the following equation to find the root of $j_{\min}=\sqrt{1-e_{\max}^2}$ \citep[e.g.,][]{liu2015}
\begin{align}
\frac{3}{8}&\times\{e_0+(j_{\min}^2-1)+(5-4j_{\min}^2)\nonumber\\
&\times\left[1-\frac{((j_{\min}^2-1)\zeta_{\min}+e_0^2\zeta_0-2j_0\cos I_0)^2}{4j_{\min}^2} \right]\nonumber\\
&-(1+4e_0^2-5e_0^2\cos^2 \omega_0
)\sin^2 I_0\}+\epsilon_{\rm GR}(j_0^{-1}-j_{\min}^{-1})\nonumber\\
&+\frac{\epsilon_{\rm Tide}}{15}\times\left(\frac{32-24j_0^2+3(1-j_0^2)^2}{8j_0^9}\right.\nonumber\\
&-\left.\frac{32-24j_{\min}^2+3(1-j_{\min}^2)^2}{8j_{\min}^9}\right)=0\,.
\label{eqn:compemax}
\end{align}

The above equation is derived in a quadrupole approximation, but has been shown to remain approximately valid even when the octupole effect is non-negligible \citep[e.g.,][]{anders2016,anders2017,liu2019}. In the previous equation, $e_0$ is the initial inner binary eccentricity, $j_0=\sqrt{1-e_0^2}$, $\zeta_{\min}=L(e=e_{\max})/L_{\rm out}$, and $\zeta_0=L(e=e_0)/L_{\rm out}$, where $L$ and $L_{\rm out}$ are the angular momenta of the inner and outer binaries, respectively. The parameters
\begin{equation} 
\epsilon_{\rm GR}=\frac{3Gm_{\rm in}^2 a_{\rm out}^3 (1-e_{\rm out}^2)^{3/2} }{c^2a_{\rm in}^4 m_{\rm out}}
\end{equation}
and\footnote{This assumes that only one of the two objects in the inner binary raises tides. If both components of the inner binary raise tides, $\epsilon_{\rm Tide}$ has a contribution from both components.}
\begin{equation}
\epsilon_{\rm Tide}=\frac{15m_{\rm in}^2 a_{\rm out}^3 (1-e_{\rm out}^2)^{3/2} k_{\rm Love,*}R_*^5}{a_{\rm in}^8 m_* m_{\rm out}}
\end{equation}
represent the relative strength of the apsidal precession due to GR and tidal bulge of the star\footnote{We do not include precession due to rotational distortion of the star, which is usually negligible.}. Here, $R_*$ and $k_{\rm Love,*}$ are the radius and the Love number of a given star, respectively. For a MS star, a good approximation is $k_{\rm Love,*}=0.028$, while for other stellar types it depends on the details of the stellar structure \citep{hut1981,kisel1998}.

\begin{figure*} 
\centering
\includegraphics[scale=0.655]{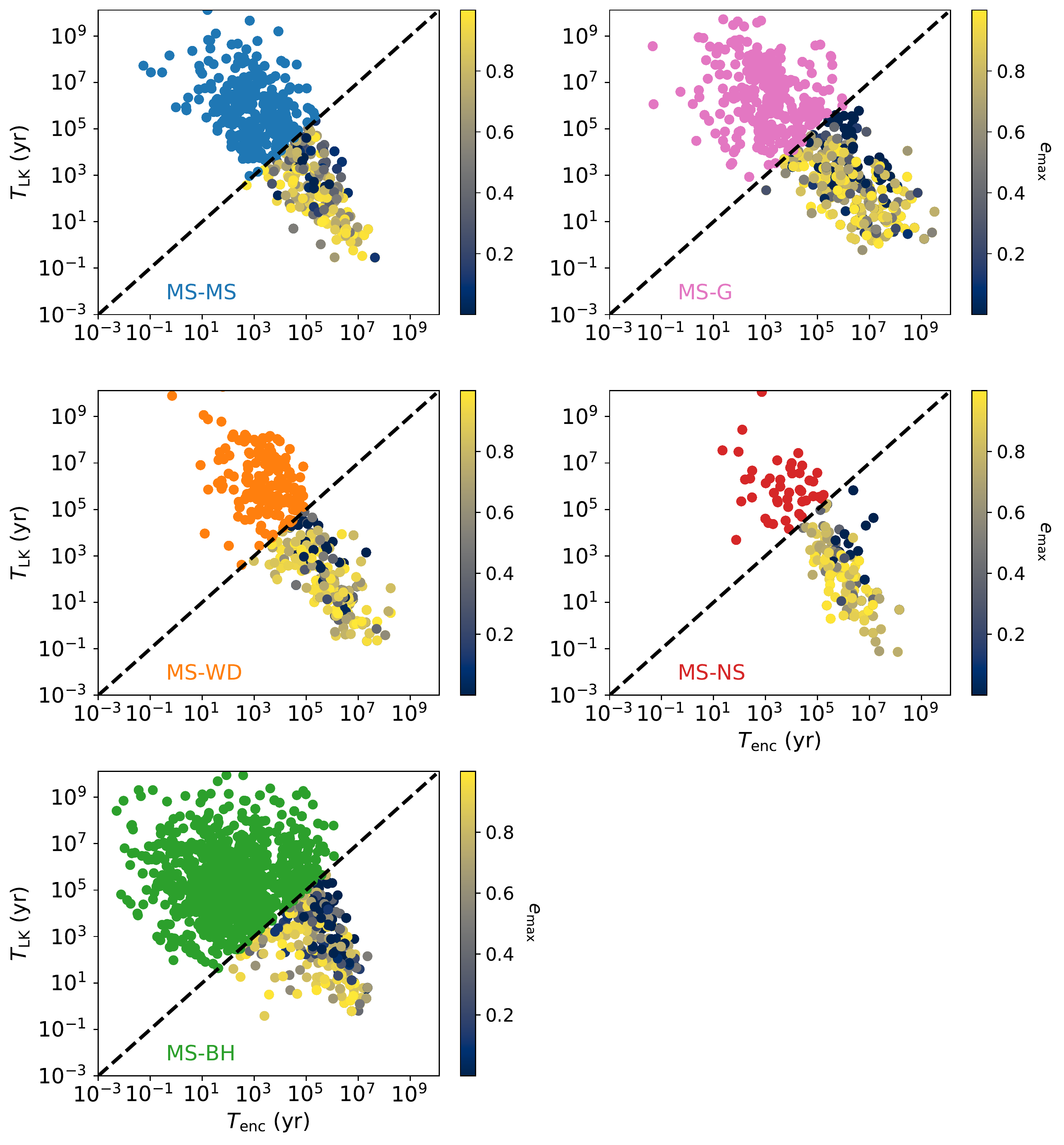}
\caption{Comparison between the LK timescale ($T_{\rm LK}$) and the encounter timescale ($T_{\rm enc}$) for triples with a MS star plus a companion in the inner binary: MS-MS (top-left panel); MS-G (top-right panel); MS-WD (center-left panel); MS-NS (center-right panel); MS-BH (bottom-left panel). Color code: maximum eccentricity attained by triples with $T_{\rm LK}<T_{\rm enc}$, computed using Eq~.\ref{eqn:compemax}.}
\label{fig:innerms}
\end{figure*}

Eccentricity excitations near unity during LK cycles can deeply alter the evolution of binary systems, the components of which would otherwise not interact if isolated from the tertiary perturber. For instance, inner binaries comprised of stars can efficiently shrink their orbit owing to efficient tides at the pericenter \citep[e.g.,][]{peretsfab2009,naozf2014,naozf2016,step2016,rose2019} or merge due to dissipation of energy via GW emission \citep[e.g.,][]{step2016,grish2018,hoang2018,Fragleipern2019,step2019}.

In a cluster's dense stellar environment, triple systems may be perturbed through encounters with other passing stars. As discussed, these encounters take place on a typical timescale $T_{\rm enc}$ (see Eq.~\ref{eqn:tenc}). Such encounters can reset the triple by altering the orbital properties significantly. In the case of soft triples, encounters with other cluster members will even tend to disrupt it, on average. Thus, unlike triples in isolation, LK cycles must occur on timescales shorter than the encounter timescale. If $T_{\rm LK}<T_{\rm enc}$, the inner binary eccentricity can reach high values and trigger the interaction, or even the merger, of the components in the inner binary. If $T_{\rm LK}>T_{\rm enc}$, LK oscillations could be suppressed by stellar encounters \citep{antoninietal2016}.

As an example, we show in Figure~\ref{fig:innerms} a comparison between the LK timescale and the encounter timescale for triples with a MS star plus a companion in the inner binary: MS-MS (top-left panel), MS-G (top-right panel), MS-WD (center-left panel), MS-NS (center-right panel), and MS-BH (bottom-left panel). In each panel, for the systems that satisfy $T_{\rm LK}<T_{\rm enc}$, we represent in color code the maximum eccentricity attained by triples, computed by using Eq.~\ref{eqn:compemax}. 

We showed in Section~\ref{sect:triporigin} that triple systems experience a recoil kick as a result of the binary--binary exchange encounter. The recoil kick can be large enough to eject the triple from the core. If not ejected from the cluster, the triple would have a new elongated orbit with pericenter in the cluster core and apocenter in the cluster outskirts. The triple would eventually sink back to the core as a result of dynamical friction (Eq.~\ref{eqn:tdf}). However, the encounter timescale of the triple would be longer than given by Eq.~\ref{eqn:tenc} since it would spend most of its orbit in regions less dense than the core. To bracket the uncertainties, we show the results of our LK analysis both in the case the encounter timescale of triples is computed using Eq.~\ref{eqn:tenc} and in the case $T_{\rm enc}$ goes to infinity (essentially corresponding to a triple ejected from the cluster environment; see Section~\ref{subsect:ejec}).

\subsection{Collision and accretion in triples with a main-sequence star, a giant, or a white dwarf in the inner binary}
\label{subsect:collaccrtrip}

During the LK evolution, the inner orbital eccentricity is excited, which can result in crossing of the Roche limit. Given a binary system with components $m_{\rm i}$ and $m_{\rm j}$, we define the dimensionless number \citep{egg83}
\begin{equation}
\mu_{\rm ji}=0.49\frac{(m_j/m_i)^{2/3}}{0.6(m_j/m_i)^{2/3}+\ln(1+(m_j/m_i)^{1/3})}\,.
\end{equation}
Thus, the Roche limit is defined as
\begin{equation}
a_{\rm Roche,ij}\equiv\frac{R_{\rm j}}{\mu_{\rm ji}}\,,
\label{eqn:rochea}
\end{equation}
where $R_{\rm j}$ is the radius of $m_{\rm j}$. The definition of $a_{\rm Roche,ji}$ is obtained with the substitutions $i\rightarrow j$ and $j\rightarrow i$. For triples that comprise of a MS star or a G star in the inner binary, we compute $e_{\rm max}$ from Eq.~\ref{eqn:compemax} and define Roche-lobe overflow to occur whenever \citep[e.g.,][]{step2019}
\begin{equation}
a(1-e_{\rm max})\le a_{\rm Roche}\,.
\end{equation}

\begin{figure*} 
\centering
\includegraphics[scale=0.45]{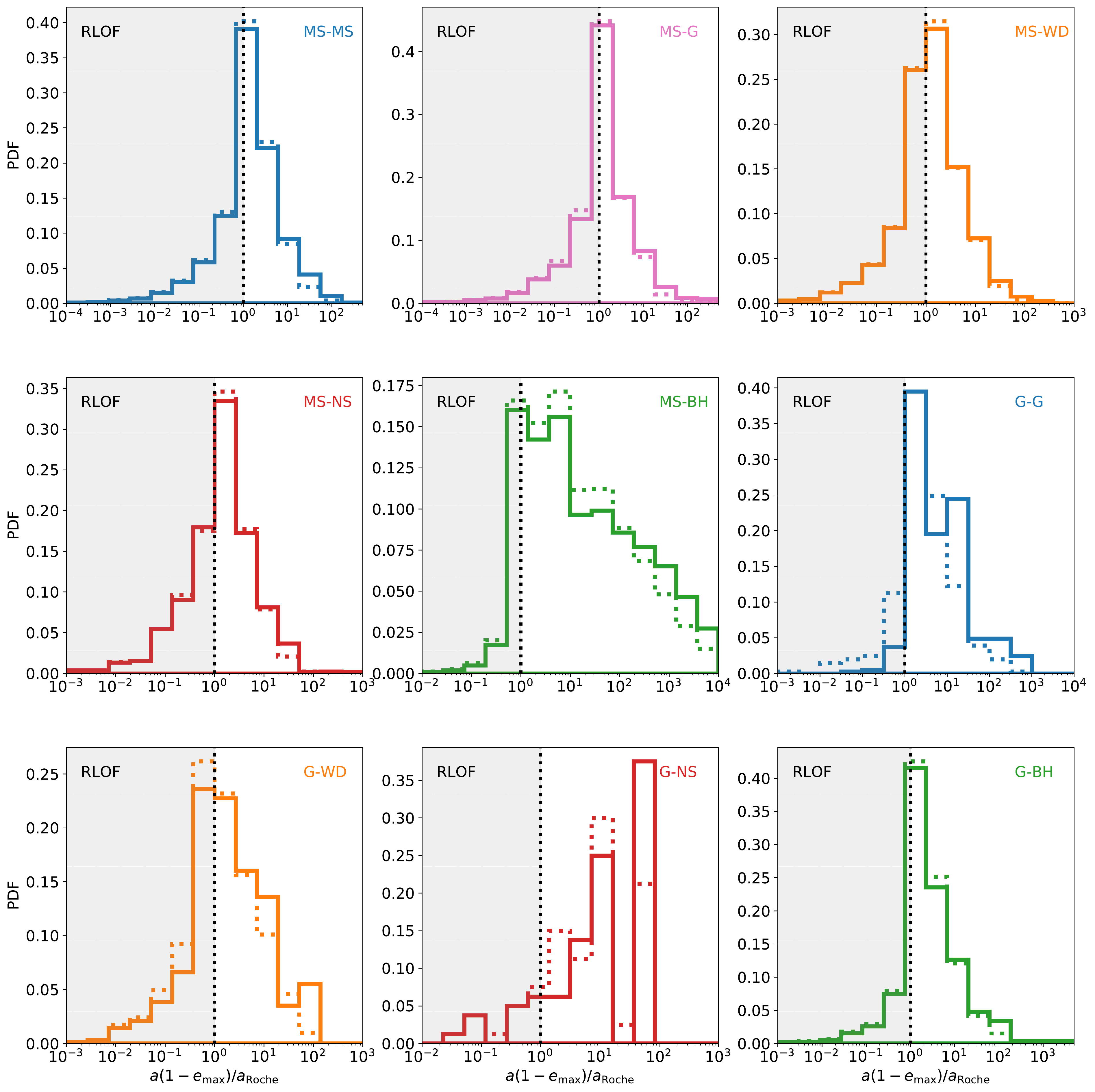}
\caption{Probability distribution function of the ratio of the inner binary's pericenter during a LK cycle to the Roche semi-major axis (Eq.~\ref{eqn:rochea}), for triples with a MS or a G star in the inner binary. The shaded area represents the region where a Roche-lobe overflow can take place ($a(1-e_{\rm max})/a_{\rm Roche}\le1$). Solid lines represent the condition $T_{\rm LK}<T_{\rm enc}$ and dotted lines represent the case where $T_{\rm enc}$ goes to infinity.}
\label{fig:emaxmsgg}
\end{figure*}

We show in Figure~\ref{fig:emaxmsgg} the probability distribution function of the ratio of the inner binary's pericenter during a LK cycle to the the Roche semi-major axis (Eq.~\ref{eqn:rochea}), for triples with a MS star or a G star in the inner binary. The shaded area represents the region where $a(1-e_{\rm max})/a_{\rm Roche}\le1$, where a Roche-lobe overflow can take place. According to the companion of the MS or G star in the inner binary of these triples, the LK cycles can produce either accretion or a physical merger. In the case of MS inner binaries, MS-MS and MS-G would likely form blue stragglers and rejuvenated giants, MS-WD would form cataclysmic variables, and MS-NS or MS-BH would give birth to X-ray binaries, millisecond pulsars, or Thorne-Zytkow objects. On the other hand, G-G mergers would form rejuvenated giants, while mergers of G with a compact object could give birth to ultracompact X-ray binaries \citep{Hurley2000,Hurley2002,Ivanova2010,naozf2016,perets2016,Kremer2018a,fragetal2019,Kremer2019c,step2019}. In Figure~\ref{fig:emaxmsgg}, we also illustrate a comparison of the systems that satisfy $a(1-e_{\rm max})/a_{\rm Roche}\le1$ when computing $T_{\rm enc}$ using Eq.~\ref{eqn:tenc} (solid line) and when $T_{\rm enc}$ goes to infinity (dotted line). We find that there is not a significant difference between using Eq.~\ref{eqn:tenc} to compute $T_{\rm enc}$ and treating $T_{\rm enc}$ as infinite, since for these systems the LK timescale is typically smaller than the encounter timescale from Eq.~\ref{eqn:tenc}.

\begin{figure} 
\centering
\includegraphics[scale=0.55]{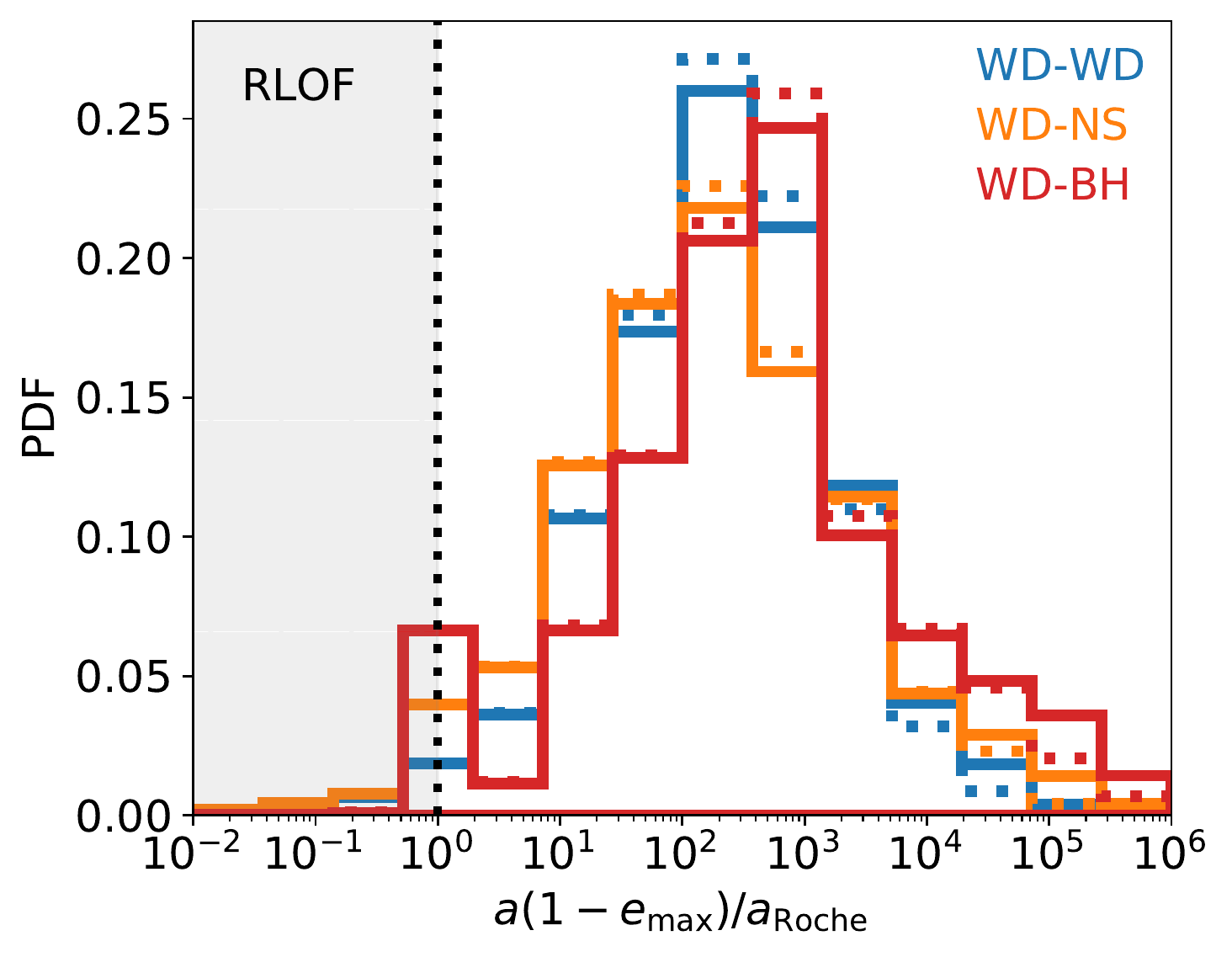}
\caption{Probability distribution function of the ratio of the inner binary's pericenter during a LK cycle and the Roche semi-major axis, for triples with a WD in the inner binary. The shaded area represents the region where a Roche-lobe overflow can take place ($a(1-e_{\rm max})/a_{\rm Roche}\le1$). Solid lines represent the condition $T_{\rm LK}<T_{\rm enc}$ and dotted lines represent the case $T_{\rm enc}$ goes to infinity.}
\label{fig:emaxwdro}
\end{figure}

We estimate that $\sim$ 35, 43, 38, 32, and 14\% of the triple systems merge with inner MS-MS, MS-G, MS-WD, MS-NS, and MS-BH binaries, respectively, while  $\sim$12, 38, 16, and 15\% of the systems merge for triples with inner G-G, G-WD, G-NS, and G-BH binaries, respectively. Assuming a GC density $\rho_{\rm{GC}}\sim 2.31\,\rm{Mpc}^{-3}$ \citep{Rodriguez2015a,RodriguezLoeb2018}, we estimate a merger rate of $\sim 10^{-1}$--$10^{-2}$ Gpc$^{-3}$ yr$^{-1}$ for these populations of triples, consistent with the previous estimated rates in cluster binaries \citep{Kremer2019c} and in field triples \citep{fragetal2019}.

In Figure~\ref{fig:emaxwdro}, we plot the probability distribution function of the ratio of the inner binary's pericenter during a LK cycle to the Roche semi-major axis, for triples with a WD in the inner binary. The shaded area represents the region where $a(1-e_{\rm max})/a_{\rm Roche}\le1$. The outcome of the accretion depends on the components of the inner binary. WD-WD mergers can lead to Type Ia SNe, while WD-NS and WD-BH mergers can lead to tidal disruption events and gamma-ray bursts \citep{Hurley2002,fryer1999,perets2016,fragmet2019,leigh2020}.

We estimate that $\sim$ 1.9, 4.6, and 4.2\% of the systems merge for triples with inner WD-WD, WD-NS, and WD-BH binaries, respectively. We find that there is no significant difference between the case where $T_{\rm enc}$ is computed using Eq.~\ref{eqn:tenc} and the case $T_{\rm enc}$ goes to infinity, since for these systems the LK timescale is typically smaller than the encounter timescale from Eq.~\ref{eqn:tenc}. Merging WDs have masses in the range $\sim 0.2\msun$--$1.4\msun$, while merging NSs and BHs have typical masses of $\sim 1.3\msun$ and $\sim 10\msun$, respectively. Assuming a GC density $\rho_{\rm{GC}}\sim 2.31\,\rm{Mpc}^{-3}$ \citep{Rodriguez2015a,RodriguezLoeb2018}, we compute a merger rate of $\sim 10^{-3}$ Gpc$^{-3}$ yr$^{-1}$, consistent with the estimated rate for this kind of merger in field triples \citep{fragmet2019}.

\subsection{Gravitational wave mergers in triples with a white dwarf, a neutron star, or a black hole in the inner binary}
\label{subsect:gwmergtrip}

For triples comprised of an inner binary with two compact objects, GW emission becomes relevant. Given a binary of components $M_1$ and $M_2$, semi-major axis $a_{12}$, and eccentricity $e_{12}$, it would merge through GW emission in isolation on a timescale \citet{Peters1964}
\begin{equation}
T_{\rm GW}=\frac{5}{256}\frac{a_{12}^4 c^5}{G^3 (M_1+M_2) M_1 M_2}(1-e_{12}^2)^{7/2}\,.
\label{eqn:tgw}
\end{equation}
When LK oscillations are relevant in a triple system, the inner binary would spend a fraction of its time $\propto (1-e^2_{\rm max})^{1/2}$ at $e\sim e_{\rm max}$, where it loses energy efficiently due to GW emission. Thus, the GW timescale would be reduced compared to a binary in isolation \citep[e.g.,][]{grish2018}
\begin{equation}
T_{\rm GW}^{\rm (red)}=\frac{5}{256}\frac{a_{12}^4 c^5}{G^3 (M_1+M_2) M_1 M_2}(1-e_{12}^2)^3\,.
\label{eqn:tgwred}
\end{equation}

\begin{figure} 
\centering
\includegraphics[scale=0.55]{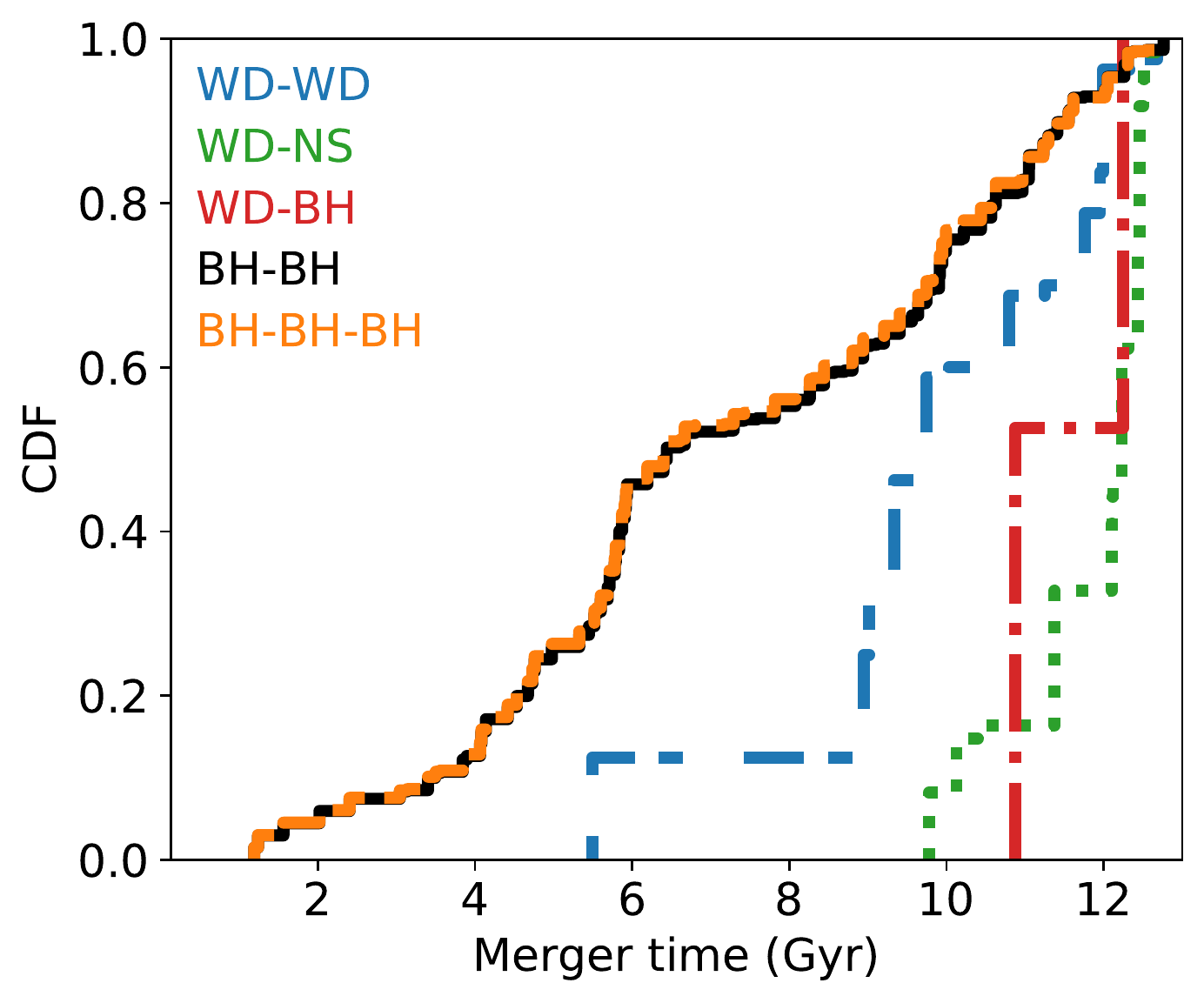}
\caption{Cumulative distribution function of the merger time ($T_{\rm f}+T_{\rm GW}^{(red)}$), for triples with an inner binary comprised of two compact objects that merge due to the LK mechanism. The LK mechanism does not produce NS-NS or BH-NS mergers in our models.}
\label{fig:emaxgwall}
\end{figure}

We show in Figure~\ref{fig:emaxgwall} the cumulative distribution function of the merger time ($T_{\rm f}+T_{\rm GW}^{(red)}$) for triples with a WD in the inner binary. If the reduced GW merger time is shorter than the LK timescale that is required to reach the maximal eccentricity, we use the secular LK time \citep{fragrish2019}. We find that $\sim$0.6, 2.5, and 0.2\% of the triples with inner WD-WD, WD-NS, and WD-BH binaries merge due to the LK mechanism within a Hubble time, respectively. We find no difference in the merger fractions when computing $T_{\rm enc}$ using Eq.~\ref{eqn:tenc} and in the case $T_{\rm enc}$ goes to infinity, respectively.

Triples with an inner WD-BH and WD-NS binary could be observed by LISA up to the point of disruption. The GW frequency at disruption is\footnote{Note that this corresponds to circular orbits, but the peak GW frequency at disruption is similar for arbitrary eccentricities to within $\sim 20\%$.} \citep{fragmet2019}
\begin{align}
f_{\rm GW} &= \frac{G^{1/2}  (M_{\rm 2}+M_{\rm WD})^{1/2} }{\pi R_{\rm t}^{3/2}}\nonumber\\
&= 0.09\,{\rm Hz}\,\left(1+\frac{M_{\rm WD}}{M_{\rm 2}}\right)M_{\rm WD,0.6\msun}^{1/2} R_{\rm WD,10^4\,{\rm km}}^{-3/2},
\label{eq:fgwwd}
\end{align}
where $R_{\rm WD}\propto M_{\rm WD}^{-1/3}$ is the WD radius and $M_2$ is the BH or NS mass\footnote{We introduced the abbreviated notation $X_{,a} = X/a$}. The total characteristic GW strain for observing the GWs for a duration $T_{\rm obs}$ averaged over binary and detector orientation is approximately \citep{Robson+2019}
\begin{align}
h_{\rm c} &= \frac{8}{\sqrt{5}} \frac{G^2}{c^4}\frac{M_{\rm 2}M_{\rm WD}}{R_t D} \left(T_{\rm obs} f_{\rm GW}\right)^{1/2}  = 2.0\times 10^{-20}\nonumber\\
&\quad\times T_{\rm obs,4\rm yr}^{0.5} D_{10\, \rm Mpc}^{-1} 
M_{\rm 2,10\msun}^{0.66} M_{\rm WD,0.6 \msun}^{1.58}
R_{\rm WD,10^4\,{\rm km}}^{-1.75}.
\end{align}

In Figure~\ref{fig:emaxgwall}, we also show the merging systems with an inner binary BH that merge due to the LK mechanism. We estimate that $\sim 0.1\%$ of triples with a binary BH as inner binary merge within a Hubble time. We find there is no significant difference between the cases where the tertiary is any kind of object (BH-BH) or a BH (BH-BH-BH), thus implying that the majority of BH mergers due to the LK mechanism take place in triple systems where all the objects are BHs. Moreover, we find no difference in the merger fractions when computing $T_{\rm enc}$ using Eq.~\ref{eqn:tenc} and in the case $T_{\rm enc}$ goes to infinity, respectively. None of the triples with a NS in the inner binary merge within a Hubble time. The reason is that triples with NSs in the inner binary are formed at late times, when most of the BHs have been ejected in the BH-burning process \citep{Kremer2019d}, as shown in Figure~\ref{fig:tinmoutm}. Therefore, triple systems likely do not contribute to the rates of NS-NS and BH-NS mergers in clusters, which remain too small to account for LIGO/Virgo observations, as shown in detail by \citet{ye2020}.

In order to estimate the local cosmological rate of BH-BH mergers in cluster triple systems, we compute the cumulative merger rate as \citep[e.g.,][]{Rodriguez2015a}
\begin{equation}
R(z) = \int_0^z \mathcal{R}(z^\prime) \frac{dV_c}{dz^\prime}(1+z^\prime)^{-1}dz^\prime\,,
\end{equation}
where $dV_c/dz$ is the comoving volume at redshift $z$ and $\mathcal{R}(z)$ is the comoving (source) merger rate. The comoving rate is given by
\begin{equation}
\label{eq:comoving_rate}
\mathcal{R}(z) = f \times \rho_{\rm{GC}} \times \frac{dN(z)}{dt}\,,
\end{equation}
where $\rho_{\rm{GC}}\sim 2.31\,\rm{Mpc}^{-3}$ \citep{Rodriguez2015a,RodriguezLoeb2018}, $f\sim 4$ is a scaling factor intended to incorporate the contribution of the cluster mass function's high-end tail not covered by our models \citep{kremer2020}, and $dN(z)/dt$ is the number of mergers per unit time at a given redshift. To estimate $dN(z)/dt$, we draw $10$ random ages for the host cluster, from the metallicity-dependent age distributions of \citet{El-Badry2018}, where the merger originated and then compute the effective merger time for each merger. We find that the merger rate for BH triples in star clusters is $\sim 0.1$ Gpc$^{-3}$ yr$^{-1}$ in the local Universe, consistent with \citet{antoninietal2016}, within the uncertainties. We leave a detailed calculation and discussion of the implications of BH mergers in triples to a companion paper (Martinez et al., submitted).

\section{Discussion and conclusions}
\label{sect:conc}

Stellar multiplicity is an omnipresent outcome of the star-formation process \citep{duc13}. More than $\sim 50$\% and $\sim 25\%$ of stars are thought to have at least one and two stellar companions, respectively. Hierarchical systems can also be formed in star clusters \citep{Fregeau2004,leighgeller2013}. In these dynamically-active environments, few-body interactions between stars and/or compact remnants can efficiently assemble hierarchical systems, primarily due to binary--binary encounters. In this process, one of the two binaries captures a star in the second wider binary, with the fourth object leaving the system. 

In this paper, we have presented for the first time the demographics of triple systems of stars and compact objects assembled in dense star clusters of various masses, concentrations, and metallicities. We have made use of the ensemble of cluster simulations presented in \citet{kremer2020}, which covers roughly the complete range of GCs observed at present day in the Milky Way. 

We have demonstrated that triples are efficiently assembled in binary--binary encounters that involve two binaries of quite different sizes. In this process, the tighter binary replaces one of the components in the wider binary. The object that is removed is then ejected, while the captured one becomes the tertiary in the newly formed triple system. During these binary--binary encounters, triple formation can lead to GW captures and mergers of stars and compact objects. We have found that a cluster typically assembles hundreds of triples with an inner BH-BH binary (of which $\sim 70$--$90\%$ have a BH as tertiary) or an inner MS-BH binary. Additionally, tens of triples with inner MS-MS and WD-BH binaries are produced. Only clusters with $r_{\rm v}\le 1$\,pc are efficient in assembling triples with inner binaries comprised of MS-WD or WD-WD pairings. Due to the BH burning process \citep{Kremer2019d}, these clusters produce $\sim 10$ times more systems with inner MS-MS binaries. We have also found that $\sim 50\%$ of the overall triple population from our simulations consists of systems where all the components are BHs.Roughly 10\% of the triples consist of an inner BH-BH binary with with a non-BH tertiary companion, while $\sim 38\%$ consist of an inner binary containing at least one MS star. Other triples constitute the remaining $\sim 2\%$ of the population.

We have shown that the initial properties of the host cluster set the typical orbital parameters and formation times of the assembled triples. Smaller and less-extended clusters form triples faster and with wider inner and outer orbits with respect to more massive and concentrated clusters. We have also found that triples whose inner binary comprises at least one BH typically form on a shorter timescale compared to other triples. This is a direct consequence of the BH-burning mechanism \citep{Kremer2019d}.

We have discussed how the LK mechanism can drive the inner binary of the formed triples to high eccentricities, whenever it takes place before the triple is dynamically reprocessed by encountering another cluster member. Some of these systems can reach sufficiently large eccentricities to form a variety of exotica, transients and GW sources, such as blue stragglers, rejuvenated giant stars, X-ray binaries, Type Ia Supernovae, Thorne-Zytkow objects, and LIGO/Virgo sources.

We have also estimated that the Milky Way's globular clusters are expected to host tens of triples with at least one luminous component at present day. Due to their high densities, only one triple star system is known to exist in GCs \citep[e.g.,][]{prodan12}. The system in question, called 4U 1820-30, is located near the centre of the GC NGC 6624 and consists of a low-mass X-ray binary with a NS primary and a WD secondary, in orbit with a period $\sim 685$ s. There is also a large luminosity variation for this system with a period of $\sim 171$ days, thought to be due to the presence of a tertiary companion \citep{grindlay88}. Another confirmed triple system in the GC M4 is made up of an inner binary comprised of a pulsar (PSR 1620-26) and a white dwarf, orbited by a substellar tertiary \citep{arz1996,rasio1995}. These systems could be naturally explained by binary--binary interactions involving planetary systems in dense stellar environments \citep{kremdoraz2019}. A few nearby open clusters are also known to have comparably high multiplicity fractions \citep[see e.g.,][for a more detailed review]{leighgeller2013}. The Hyades \citep{patience98}, Pleiades \citep{mermilliod92,bouvier97} and Praesepe \citep{mermilliod99,bouvier01} have binary fractions of, respectively, $35\%$, $34\%$ and $40\%$, and triple fractions of, respectively, $6\%$, $3\%$ and $6\%$. Notably, the open cluster Taurus-Auriga appears to have a multiplicity fraction higher than the field. \citet{kraus11} performed a high-resolution imaging study to characterize the multiple-star populations in Taurus-Auriga. They found that $\sim$ $2/3$--$3/4$ of all targets are multiples composed of at least two stars. Therefore, only $\sim$ $1/4$--$1/3$ of their objects are single stars.

Triple and hierarchical systems constitute a fundamental building block for many astrophysical phenomena, which are difficult to achieve with standard binary evolution \citep{naoz2016}. While current observations improve and provide unprecedented data on the galactic field population of triples, little is known on the triple population that can be assembled in dense star clusters. Upcoming instruments, such as LSST and JWST, may shed light on this population, which critically depends on the initial properties of the parent cluster and its evolutionary paths. In particular, a crucial role can be played by the primordial binary fraction and the mass-ratio distributions of low- and high-mass stars. We leave to a future study further investigation of how triple formation and demographics depends on these parameters (Fragione et al. in prep.). While our current understanding of hierarchies in dense star clusters is still limited, the future of triple systems appears bright.

\section*{Acknowledgements}

We are grateful to Nathan Leigh, Michael Shara, Rosanne di Stefano, and Selma de Mink for useful discussions. Our work was supported by NSF Grant AST-1716762. Computations were supported in part through the resources and staff contributions provided for the Quest high performance computing facility at Northwestern University, which is jointly supported by the Office of the Provost, the Office for Research, and Northwestern University Information Technology. This work also used computing resources at CIERA funded by NSF Grant PHY-1726951 and computing resources provided by Northwestern University and the Center for Interdisciplinary Exploration and Research in Astrophysics (CIERA). GF acknowledges support from a CIERA postdoctoral fellowship at Northwestern University. SC acknowledges support from the Department of Atomic Energy, Government of India, under project no. 12-R\&D-TFR-5.02-0200. SN acknowledges the partial support of NASA grant No.~80NSSC19K0321 and No.~80NSSC20K0505, also thanks Howard and Astrid Preston for their generous support.

\appendix

\setcounter{table}{0}
\renewcommand{\thetable}{A\arabic{table}}

\section{Triple systems formed in cluster simulations}

\startlongtable
\begin{deluxetable*}{ccccc|ccccc|cccc}
\tabletypesize{\scriptsize}
\tablewidth{0pt}
\tablecaption{Initial cluster parameters and number of different triples formed. Triples with a main-sequence or a giant plus a companion in the inner binary. \label{table:models}}
\tablehead{
\colhead{} & \colhead{$r_{\rm v}$ (pc)} & \colhead{$r_{\rm g}$ (kpc)} & \colhead{$Z$} & \colhead{$N$} & \colhead{MS-MS} & \colhead{MS-G} & \colhead{MS-WD} & \colhead{MS-NS} & \colhead{MS-BH} & \colhead{G-G} & \colhead{G-WD} & \colhead{G-NS} & \colhead{G-BH}
}
\startdata
\hline\hline
1 & 0.5 & 2 & 0.0002 & $2\times 10^5$ & 150 & 16 & 17 & 0 & 314 & 0 & 0 & 0 & 6 \\
2 & 0.5 & 2 & 0.0002 & $4\times 10^5$ & 326 & 12 & 67 & 0 & 1192 & 1 & 1 & 0 & 6 \\
3 & 0.5 & 2 & 0.0002 & $8\times 10^5$ & 180 & 1 & 33 & 13 & 312 & 0 & 0 & 0 & 4 \\
4$^\dagger$ & 0.5 & 2 & 0.0002 & $1.6\times 10^6$ & 24 & 0 & 0 & 0 & 0 & 0 & 0 & 0 & 0 \\
\hline
5 & 0.5 & 2 & 0.002 & $2\times 10^5$ & 314 & 21 & 21 & 1 & 553 & 0 & 2 & 0 & 0 \\
6 & 0.5 & 2 & 0.002 & $4\times 10^5$ & 261 & 6 & 78 & 6 & 330 & 0 & 2 & 0 & 5 \\
7 & 0.5 & 2 & 0.002 & $8\times 10^5$ & 251 & 7 & 66 & 7 & 259 & 0 & 1 & 2 & 0 \\
8 & 0.5 & 2 & 0.002 & $1.6\times 10^6$ & 201 & 3 & 2 & 0 & 154 & 0 & 0 & 0 & 3 \\
\hline
9 & 0.5 & 2 & 0.02 & $2\times 10^5$ & 283 & 37 & 10 & 4 & 40 & 14 & 1 & 0 & 1 \\
10 & 0.5 & 2 & 0.02 & $4\times 10^5$ & 298 & 23 & 50 & 5 & 51 & 3 & 10 & 0 & 0 \\
11 & 0.5 & 2 & 0.02 & $8\times 10^5$ & 342 & 16 & 51 & 9 & 101 & 0 & 0 & 1 & 8 \\
12 & 0.5 & 2 & 0.02 & $1.6\times 10^6$ & 412 & 10 & 81 & 10 & 109 & 0 & 0 & 0 & 2 \\
\hline
13 & 0.5 & 8 & 0.0002 & $2\times 10^5$ & 281 & 44 & 44 & 2 & 664 & 3 & 2 & 0 & 16 \\
14 & 0.5 & 8 & 0.0002 & $4\times 10^5$ & 248 & 2 & 37 & 4 & 866 & 0 & 3 & 0 & 5 \\
15 & 0.5 & 8 & 0.0002 & $8\times 10^5$ & 184 & 1 & 41 & 6 & 258 & 0 & 1 & 0 & 1 \\
16$^\dagger$ & 0.5 & 8 & 0.0002 & $1.6\times 10^6$ & 35 & 0 & 0 & 0 & 0 & 0 & 0 & 0 & 0 \\
\hline
17 & 0.5 & 8 & 0.002 & $2\times 10^5$ & 293 & 10 & 66 & 6 & 358 & 0 & 1 & 0 & 5 \\
18 & 0.5 & 8 & 0.002 & $4\times 10^5$ & 247 & 4 & 53 & 5 & 350 & 0 & 0 & 0 & 2 \\
19 & 0.5 & 8 & 0.002 & $8\times 10^5$ & 237 & 7 & 34 & 8 & 230 & 0 & 0 & 0 & 1 \\
20 & 0.5 & 8 & 0.002 & $1.6\times 10^6$ & 190 & 1 & 1 & 0 & 97 & 0 & 0 & 0 & 1 \\
\hline
21 & 0.5 & 8 & 0.02 & $2\times 10^5$ & 221 & 21 & 38 & 1 & 134 & 3 & 1 & 0 & 3 \\
22 & 0.5 & 8 & 0.02 & $4\times 10^5$ & 279 & 17 & 38 & 2 & 100 & 0 & 7 & 3 & 9 \\
23 & 0.5 & 8 & 0.02 & $8\times 10^5$ & 283 & 10 & 53 & 3 & 122 & 0 & 2 & 0 & 2 \\
24 & 0.5 & 8 & 0.02 & $1.6\times 10^6$ & 349 & 7 & 60 & 4 & 142 & 0 & 2 & 0 & 6 \\
\hline
25 & 0.5 & 20 & 0.0002 & $2\times 10^5$ & 232 & 6 & 70 & 1 & 600 & 0 & 1 & 0 & 1 \\
26 & 0.5 & 20 & 0.0002 & $4\times 10^5$ & 294 & 2 & 21 & 3 & 623 & 0 & 0 & 0 & 3 \\
27 & 0.5 & 20 & 0.0002 & $8\times 10^5$ & 168 & 3 & 39 & 8 & 308 & 0 & 0 & 0 & 1 \\
28$^\dagger$ & 0.5 & 20 & 0.0002 & $1.6\times 10^6$ & 38 & 0 & 0 & 0 & 0 & 0 & 0 & 0 & 0 \\
\hline
29 & 0.5 & 20 & 0.002 & $2\times 10^5$ & 298 & 17 & 86 & 3 & 463 & 4 & 5 & 0 & 3 \\
30 & 0.5 & 20 & 0.002 & $4\times 10^5$ & 272 & 6 & 44 & 2 & 551 & 0 & 2 & 0 & 1 \\
31 & 0.5 & 20 & 0.002 & $8\times 10^5$ & 187 & 3 & 29 & 1 & 180 & 0 & 1 & 0 & 0 \\
32 & 0.5 & 20 & 0.002 & $1.6\times 10^6$ & 132 & 5 & 1 & 0 & 160 & 0 & 0 & 0 & 0 \\
\hline
33 & 0.5 & 20 & 0.02 & $2\times 10^5$ & 301 & 25 & 38 & 2 & 84 & 1 & 3 & 0 & 2 \\
34 & 0.5 & 20 & 0.02 & $4\times 10^5$ & 277 & 5 & 24 & 2 & 59 & 0 & 0 & 0 & 6 \\
35 & 0.5 & 20 & 0.02 & $8\times 10^5$ & 291 & 13 & 46 & 5 & 100 & 0 & 3 & 0 & 3 \\
36 & 0.5 & 20 & 0.02 & $1.6\times 10^6$ & 360 & 8 & 62 & 8 & 103 & 0 & 3 & 0 & 2 \\
\hline
37 & 1 & 2 & 0.0002 & $2\times 10^5$ & 70 & 2 & 38 & 2 & 259 & 0 & 1 & 0 & 4 \\
38 & 1 & 2 & 0.0002 & $4\times 10^5$ & 81 & 1 & 61 & 3 & 209 & 0 & 3 & 0 & 1 \\
39 & 1 & 2 & 0.0002 & $8\times 10^5$ & 17 & 1 & 2 & 0 & 159 & 0 & 0 & 0 & 0 \\
40 & 1 & 2 & 0.0002 & $1.6\times 10^6$ & 21 & 0 & 0 & 0 & 85 & 0 & 0 & 0 & 1 \\
\hline
41 & 1 & 2 & 0.002 & $2\times 10^5$ & 85 & 1 & 34 & 0 & 252 & 0 & 3 & 0 & 3 \\
42 & 1 & 2 & 0.002 & $4\times 10^5$ & 92 & 4 & 57 & 2 & 370 & 0 & 2 & 0 & 0 \\
43 & 1 & 2 & 0.002 & $8\times 10^5$ & 17 & 0 & 0 & 0 & 227 & 0 & 0 & 0 & 0 \\
44 & 1 & 2 & 0.002 & $1.6\times 10^6$ & 8 & 0 & 0 & 0 & 87 & 0 & 0 & 0 & 0 \\
\hline
45 & 1 & 2 & 0.02 & $2\times 10^5$ & 127 & 11 & 34 & 1 & 32 & 5 & 4 & 0 & 9 \\
46 & 1 & 2 & 0.02 & $4\times 10^5$ & 157 & 3 & 36 & 1 & 105 & 0 & 4 & 1 & 11 \\
47 & 1 & 2 & 0.02 & $8\times 10^5$ & 182 & 15 & 44 & 1 & 62 & 0 & 2 & 0 & 3 \\
48 & 1 & 2 & 0.02 & $1.6\times 10^6$ & 97 & 4 & 6 & 0 & 107 & 0 & 0 & 0 & 3 \\
\hline
49 & 1 & 8 & 0.0002 & $2\times 10^5$ & 97 & 2 & 48 & 1 & 528 & 0 & 2 & 0 & 1 \\
50 & 1 & 8 & 0.0002 & $4\times 10^5$ & 50 & 0 & 35 & 5 & 332 & 0 & 0 & 0 & 0 \\
51 & 1 & 8 & 0.0002 & $8\times 10^5$ & 15 & 0 & 0 & 0 & 150 & 0 & 0 & 0 & 0 \\
52 & 1 & 8 & 0.0002 & $1.6\times 10^6$ & 17 & 2 & 1 & 0 & 88 & 0 & 0 & 0 & 0 \\
\hline
53 & 1 & 8 & 0.002 & $2\times 10^5$ & 119 & 4 & 41 & 0 & 396 & 0 & 1 & 0 & 2 \\
54 & 1 & 8 & 0.002 & $4\times 10^5$ & 30 & 1 & 2 & 0 & 163 & 0 & 1 & 0 & 1 \\
55 & 1 & 8 & 0.002 & $8\times 10^5$ & 17 & 0 & 0 & 0 & 137 & 0 & 0 & 0 & 0 \\
56 & 1 & 8 & 0.002 & $1.6\times 10^6$ & 15 & 0 & 1 & 0 & 141 & 0 & 0 & 0 & 0 \\
\hline
57 & 1 & 8 & 0.02 & $2\times 10^5$ & 142 & 16 & 45 & 0 & 76 & 3 & 2 & 0 & 8 \\
58 & 1 & 8 & 0.02 & $4\times 10^5$ & 158 & 18 & 46 & 1 & 63 & 0 & 2 & 0 & 4 \\
59 & 1 & 8 & 0.02 & $8\times 10^5$ & 159 & 11 & 20 & 0 & 65 & 0 & 1 & 0 & 3 \\
60 & 1 & 8 & 0.02 & $1.6\times 10^6$ & 88 & 3 & 2 & 0 & 78 & 0 & 0 & 0 & 3 \\
\hline
61 & 1 & 20 & 0.0002 & $2\times 10^5$ & 72 & 2 & 53 & 3 & 309 & 0 & 1 & 0 & 1 \\
62 & 1 & 20 & 0.0002 & $4\times 10^5$ & 82 & 2 & 59 & 9 & 532 & 0 & 0 & 0 & 2 \\
63 & 1 & 20 & 0.0002 & $8\times 10^5$ & 15 & 1 & 0 & 0 & 135 & 0 & 0 & 0 & 0 \\
64 & 1 & 20 & 0.0002 & $1.6\times 10^6$ & 10 & 0 & 0 & 0 & 130 & 0 & 0 & 0 & 0 \\
\hline
65 & 1 & 20 & 0.002 & $2\times 10^5$ & 104 & 7 & 30 & 0 & 566 & 0 & 1 & 0 & 3 \\
66 & 1 & 20 & 0.002 & $4\times 10^5$ & 51 & 1 & 12 & 1 & 331 & 0 & 0 & 0 & 4 \\
67 & 1 & 20 & 0.002 & $8\times 10^5$ & 14 & 0 & 1 & 0 & 294 & 0 & 0 & 0 & 1 \\
68 & 1 & 20 & 0.002 & $1.6\times 10^6$ & 12 & 1 & 0 & 0 & 91 & 0 & 0 & 0 & 0 \\
\hline
69 & 1 & 20 & 0.02 & $2\times 10^5$ & 120 & 29 & 57 & 0 & 90 & 2 & 3 & 0 & 2 \\
70 & 1 & 20 & 0.02 & $4\times 10^5$ & 123 & 5 & 29 & 1 & 99 & 0 & 0 & 0 & 11 \\
71 & 1 & 20 & 0.02 & $8\times 10^5$ & 120 & 7 & 10 & 1 & 86 & 0 & 0 & 1 & 2 \\
72 & 1 & 20 & 0.02 & $1.6\times 10^6$ & 89 & 0 & 2 & 0 & 40 & 0 & 0 & 0 & 0 \\
\hline
73 & 2 & 2 & 0.0002 & $2\times 10^5$ & 46 & 1 & 15 & 0 & 559 & 0 & 0 & 0 & 2 \\
74 & 2 & 2 & 0.0002 & $4\times 10^5$ & 34 & 0 & 20 & 0 & 269 & 0 & 0 & 0 & 1 \\
75 & 2 & 2 & 0.0002 & $8\times 10^5$ & 9 & 0 & 0 & 0 & 143 & 0 & 0 & 0 & 0 \\
76 & 2 & 2 & 0.0002 & $1.6\times 10^6$ & 7 & 0 & 0 & 0 & 67 & 0 & 0 & 0 & 0 \\
\hline
77 & 2 & 2 & 0.002 & $2\times 10^5$ & 34 & 3 & 16 & 0 & 218 & 0 & 1 & 0 & 0 \\
78 & 2 & 2 & 0.002 & $4\times 10^5$ & 34 & 1 & 13 & 0 & 292 & 0 & 0 & 0 & 0 \\
79 & 2 & 2 & 0.002 & $8\times 10^5$ & 8 & 0 & 1 & 0 & 224 & 0 & 0 & 0 & 0 \\
80 & 2 & 2 & 0.002 & $1.6\times 10^6$ & 8 & 0 & 0 & 0 & 72 & 0 & 0 & 0 & 2 \\
\hline
81 & 2 & 2 & 0.02 & $2\times 10^5$ & 35 & 8 & 4 & 0 & 29 & 0 & 0 & 0 & 2 \\
82 & 2 & 2 & 0.02 & $4\times 10^5$ & 34 & 5 & 9 & 0 & 26 & 0 & 0 & 0 & 1 \\
83 & 2 & 2 & 0.02 & $8\times 10^5$ & 30 & 0 & 0 & 0 & 18 & 0 & 0 & 0 & 3 \\
84 & 2 & 2 & 0.02 & $1.6\times 10^6$ & 37 & 0 & 0 & 0 & 21 & 0 & 0 & 0 & 0 \\
\hline
85 & 2 & 8 & 0.0002 & $2\times 10^5$ & 2 & 0 & 6 & 0 & 300 & 0 & 0 & 0 & 0 \\
86 & 2 & 8 & 0.0002 & $4\times 10^5$ & 9 & 0 & 2 & 0 & 226 & 0 & 0 & 0 & 0 \\
87 & 2 & 8 & 0.0002 & $8\times 10^5$ & 8 & 0 & 0 & 0 & 65 & 0 & 0 & 0 & 1 \\
88 & 2 & 8 & 0.0002 & $1.6\times 10^6$ & 8 & 0 & 0 & 0 & 46 & 0 & 0 & 0 & 0 \\
\hline
89 & 2 & 8 & 0.002 & $2\times 10^5$ & 64 & 6 & 48 & 0 & 390 & 0 & 3 & 0 & 4 \\
90 & 2 & 8 & 0.002 & $4\times 10^5$ & 8 & 0 & 0 & 0 & 307 & 0 & 0 & 0 & 0 \\
91 & 2 & 8 & 0.002 & $8\times 10^5$ & 7 & 0 & 0 & 0 & 105 & 0 & 0 & 0 & 0 \\
92 & 2 & 8 & 0.002 & $1.6\times 10^6$ & 5 & 0 & 0 & 0 & 81 & 0 & 0 & 0 & 0 \\
\hline
93 & 2 & 8 & 0.02 & $2\times 10^5$ & 27 & 0 & 4 & 0 & 20 & 0 & 0 & 0 & 1 \\
94 & 2 & 8 & 0.02 & $4\times 10^5$ & 23 & 0 & 5 & 0 & 20 & 0 & 0 & 0 & 2 \\
95 & 2 & 8 & 0.02 & $8\times 10^5$ & 16 & 1 & 1 & 0 & 26 & 1 & 0 & 0 & 2 \\
96 & 2 & 8 & 0.02 & $1.6\times 10^6$ & 35 & 1 & 0 & 0 & 20 & 0 & 0 & 0 & 0 \\
\hline
97 & 2 & 20 & 0.0002 & $2\times 10^5$ & 14 & 0 & 6 & 0 & 324 & 0 & 0 & 0 & 1 \\
98 & 2 & 20 & 0.0002 & $4\times 10^5$ & 15 & 0 & 2 & 1 & 212 & 0 & 0 & 0 & 0 \\
99 & 2 & 20 & 0.0002 & $8\times 10^5$ & 2 & 0 & 0 & 0 & 124 & 0 & 0 & 0 & 3 \\
100 & 2 & 20 & 0.0002 & $1.6\times 10^6$ & 9 & 0 & 0 & 0 & 47 & 0 & 0 & 0 & 0 \\
\hline
101 & 2 & 20 & 0.002 & $2\times 10^5$ & 16 & 0 & 0 & 0 & 231 & 0 & 0 & 0 & 0 \\
102 & 2 & 20 & 0.002 & $4\times 10^5$ & 11 & 0 & 1 & 0 & 186 & 0 & 0 & 0 & 0 \\
103 & 2 & 20 & 0.002 & $8\times 10^5$ & 2 & 0 & 1 & 0 & 126 & 0 & 0 & 0 & 0 \\
104 & 2 & 20 & 0.002 & $1.6\times 10^6$ & 10 & 1 & 0 & 0 & 70 & 0 & 0 & 0 & 1 \\
\hline
105 & 2 & 20 & 0.02 & $2\times 10^5$ & 29 & 4 & 5 & 0 & 16 & 0 & 0 & 0 & 0 \\
106 & 2 & 20 & 0.02 & $4\times 10^5$ & 23 & 1 & 0 & 0 & 37 & 0 & 0 & 0 & 1 \\
107 & 2 & 20 & 0.02 & $8\times 10^5$ & 19 & 0 & 1 & 0 & 20 & 0 & 0 & 0 & 2 \\
108 & 2 & 20 & 0.02 & $1.6\times 10^6$ & 26 & 1 & 0 & 0 & 75 & 1 & 0 & 0 & 0 \\
\hline
109 & 4 & 2 & 0.0002 & $2\times 10^5$ & 1 & 0 & 0 & 0 & 236 & 0 & 0 & 0 & 0 \\
110 & 4 & 2 & 0.0002 & $4\times 10^5$ & 1 & 0 & 0 & 0 & 167 & 0 & 0 & 0 & 0 \\
111 & 4 & 2 & 0.0002 & $8\times 10^5$ & 2 & 0 & 0 & 0 & 60 & 0 & 0 & 0 & 0 \\
112 & 4 & 2 & 0.0002 & $1.6\times 10^6$ & 5 & 0 & 0 & 0 & 265 & 0 & 0 & 0 & 0 \\
\hline
113 & 4 & 2 & 0.002 & $2\times 10^5$ & 1 & 1 & 0 & 0 & 265 & 0 & 0 & 0 & 0 \\
114 & 4 & 2 & 0.002 & $4\times 10^5$ & 3 & 0 & 0 & 0 & 198 & 0 & 0 & 0 & 0 \\
115 & 4 & 2 & 0.002 & $8\times 10^5$ & 0 & 0 & 0 & 0 & 167 & 0 & 0 & 0 & 0 \\
116 & 4 & 2 & 0.002 & $1.6\times 10^6$ & 4 & 0 & 0 & 0 & 104 & 0 & 0 & 0 & 0 \\
\hline
117 & 4 & 2 & 0.02 & $2\times 10^5$ & 1 & 0 & 0 & 0 & 3 & 0 & 0 & 0 & 0 \\
118 & 4 & 2 & 0.02 & $4\times 10^5$ & 5 & 0 & 2 & 0 & 11 & 0 & 0 & 0 & 1 \\
119 & 4 & 2 & 0.02 & $8\times 10^5$ & 11 & 0 & 0 & 0 & 11 & 0 & 0 & 0 & 1 \\
120 & 4 & 2 & 0.02 & $1.6\times 10^6$ & 17 & 0 & 2 & 0 & 10 & 0 & 0 & 0 & 0 \\
\hline
121 & 4 & 8 & 0.0002 & $2\times 10^5$ & 4 & 0 & 0 & 0 & 554 & 0 & 0 & 0 & 7 \\
122 & 4 & 8 & 0.0002 & $4\times 10^5$ & 2 & 0 & 0 & 0 & 205 & 0 & 0 & 0 & 0 \\
123 & 4 & 8 & 0.0002 & $8\times 10^5$ & 2 & 0 & 0 & 0 & 64 & 0 & 0 & 0 & 0 \\
124 & 4 & 8 & 0.0002 & $1.6\times 10^6$ & 2 & 0 & 0 & 0 & 36 & 0 & 0 & 0 & 0 \\
\hline
125 & 4 & 8 & 0.002 & $2\times 10^5$ & 4 & 0 & 1 & 0 & 280 & 0 & 0 & 0 & 0 \\
126 & 4 & 8 & 0.002 & $4\times 10^5$ & 6 & 0 & 1 & 0 & 211 & 0 & 0 & 0 & 0 \\
127 & 4 & 8 & 0.002 & $8\times 10^5$ & 3 & 0 & 0 & 0 & 178 & 0 & 0 & 0 & 0 \\
128 & 4 & 8 & 0.002 & $1.6\times 10^6$ & 6 & 0 & 0 & 0 & 125 & 0 & 0 & 0 & 0 \\
\hline
129 & 4 & 8 & 0.02 & $2\times 10^5$ & 7 & 0 & 2 & 0 & 13 & 0 & 0 & 0 & 0 \\
130 & 4 & 8 & 0.02 & $4\times 10^5$ & 7 & 0 & 7 & 0 & 10 & 0 & 0 & 0 & 0 \\
131 & 4 & 8 & 0.02 & $8\times 10^5$ & 7 & 0 & 2 & 0 & 11 & 0 & 0 & 0 & 1 \\
132 & 4 & 8 & 0.02 & $1.6\times 10^6$ & 14 & 1 & 0 & 0 & 10 & 0 & 0 & 0 & 2 \\
\hline
133 & 4 & 20 & 0.0002 & $2\times 10^5$ & 1 & 0 & 0 & 0 & 231 & 0 & 0 & 0 & 0 \\
134 & 4 & 20 & 0.0002 & $4\times 10^5$ & 2 & 0 & 2 & 0 & 168 & 0 & 0 & 0 & 2 \\
135 & 4 & 20 & 0.0002 & $8\times 10^5$ & 2 & 0 & 0 & 0 & 83 & 0 & 0 & 0 & 0 \\
136 & 4 & 20 & 0.0002 & $1.6\times 10^6$ & 6 & 0 & 0 & 0 & 34 & 0 & 0 & 0 & 0 \\
\hline
137 & 4 & 20 & 0.002 & $2\times 10^5$ & 6 & 1 & 2 & 0 & 250 & 0 & 0 & 0 & 1 \\
138 & 4 & 20 & 0.002 & $4\times 10^5$ & 4 & 0 & 0 & 0 & 268 & 0 & 0 & 0 & 1 \\
139 & 4 & 20 & 0.002 & $8\times 10^5$ & 4 & 0 & 0 & 0 & 175 & 0 & 0 & 0 & 0 \\
140 & 4 & 20 & 0.002 & $1.6\times 10^6$ & 4 & 0 & 0 & 0 & 59 & 0 & 0 & 0 & 0 \\
\hline
141 & 4 & 20 & 0.02 & $2\times 10^5$ & 5 & 0 & 1 & 0 & 10 & 0 & 0 & 0 & 1 \\
142 & 4 & 20 & 0.02 & $4\times 10^5$ & 10 & 0 & 0 & 0 & 5 & 0 & 0 & 0 & 0 \\
143 & 4 & 20 & 0.02 & $8\times 10^5$ & 10 & 1 & 1 & 0 & 6 & 0 & 0 & 0 & 0 \\
144 & 4 & 20 & 0.02 & $1.6\times 10^6$ & 13 & 0 & 1 & 0 & 8 & 0 & 0 & 0 & 0 \\
\hline
145 & 1 & 20 & 0.0002 & $3.2\times 10^6$ & 14 & 0 & 0 & 0 & 26 & 0 & 0 & 0 & 0 \\
146 & 2 & 20 & 0.0002 & $3.2\times 10^6$ & 11 & 0 & 0 & 0 & 37 & 0 & 0 & 0 & 0 \\
147 & 1 & 20 & 0.02 & $3.2\times 10^6$ & 49 & 0 & 0 & 0 & 29 & 0 & 0 & 0 & 0 \\
148 & 2 & 20 & 0.02 & $3.2\times 10^6$ & 44 & 1 & 3 & 0 & 31 & 0 & 0 & 0 & 1 \\
\hline
\enddata
\tablecomments{Models marked with a dagger ($^\dagger$) indicates the model was stopped due to onset of collisional runaway \citep[see][for details]{kremer2020}.}
\end{deluxetable*}

\startlongtable
\begin{deluxetable*}{ccccc|ccc|cc|cc}
\tabletypesize{\scriptsize}
\tablewidth{0pt}
\tablecaption{Initial cluster parameters and number of different triples formed. Triples with a white dwarf, neutron star, or a black hole plus a companion in the inner binary. \label{table:models2}}
\tablehead{
\colhead{} & \colhead{$r_{\rm v}$ (pc)} & \colhead{$r_{\rm g}$ (kpc)} & \colhead{$Z$} & \colhead{$N$} & \colhead{WD-WD} & \colhead{WD-NS} & \colhead{WD-BH} & \colhead{NS-NS} & \colhead{NS-BH} & \colhead{BH-BH} & \colhead{BH-BH-BH}
}
\startdata
\hline\hline
1 & 0.5 & 2 & 0.0002 & $2\times 10^5$ & 2 & 0 & 11 & 0 & 0 & 265 & 170 \\
2 & 0.5 & 2 & 0.0002 & $4\times 10^5$ & 18 & 11 & 3 & 0 & 0 & 562 & 301 \\
3 & 0.5 & 2 & 0.0002 & $8\times 10^5$ & 42 & 16 & 27 & 4 & 0 & 755 & 635 \\
4$^\dagger$ & 0.5 & 2 & 0.0002 & $1.6\times 10^6$ & 0 & 0 & 0 & 0 & 0 & 0 & 0 \\
\hline
5 & 0.5 & 2 & 0.002 & $2\times 10^5$ & 6 & 0 & 7 & 0 & 0 & 263 & 232 \\
6 & 0.5 & 2 & 0.002 & $4\times 10^5$ & 50 & 3 & 10 & 0 & 1 & 422 & 369 \\
7 & 0.5 & 2 & 0.002 & $8\times 10^5$ & 30 & 15 & 8 & 1 & 0 & 662 & 610 \\
8 & 0.5 & 2 & 0.002 & $1.6\times 10^6$ & 0 & 0 & 3 & 0 & 0 & 973 & 949 \\
\hline
9 & 0.5 & 2 & 0.02 & $2\times 10^5$ & 1 & 0 & 0 & 0 & 0 & 188 & 152 \\
10 & 0.5 & 2 & 0.02 & $4\times 10^5$ & 25 & 8 & 1 & 0 & 0 & 263 & 234 \\
11 & 0.5 & 2 & 0.02 & $8\times 10^5$ & 18 & 11 & 8 & 0 & 1 & 291 & 271 \\
12 & 0.5 & 2 & 0.02 & $1.6\times 10^6$ & 18 & 4 & 17 & 1 & 0 & 438 & 412 \\
\hline
13 & 0.5 & 8 & 0.0002 & $2\times 10^5$ & 20 & 12 & 5 & 0 & 2 & 246 & 125 \\
14 & 0.5 & 8 & 0.0002 & $4\times 10^5$ & 125 & 31 & 75 & 3 & 0 & 692 & 467 \\
15 & 0.5 & 8 & 0.0002 & $8\times 10^5$ & 34 & 24 & 16 & 3 & 6 & 797 & 694 \\
16$^\dagger$ & 0.5 & 8 & 0.0002 & $1.6\times 10^6$ & 0 & 0 & 0 & 0 & 0 & 0 & 0 \\
\hline
17 & 0.5 & 8 & 0.002 & $2\times 10^5$ & 41 & 1 & 1 & 0 & 0 & 267 & 236 \\
18 & 0.5 & 8 & 0.002 & $4\times 10^5$ & 16 & 0 & 9 & 0 & 0 & 479 & 438 \\
19 & 0.5 & 8 & 0.002 & $8\times 10^5$ & 24 & 4 & 12 & 2 & 0 & 555 & 521 \\
20 & 0.5 & 8 & 0.002 & $1.6\times 10^6$ & 0 & 0 & 0 & 0 & 0 & 898 & 867 \\
\hline
21 & 0.5 & 8 & 0.02 & $2\times 10^5$ & 54 & 3 & 3 & 0 & 0 & 82 & 72 \\
22 & 0.5 & 8 & 0.02 & $4\times 10^5$ & 23 & 1 & 20 & 0 & 0 & 150 & 141 \\
23 & 0.5 & 8 & 0.02 & $8\times 10^5$ & 11 & 4 & 41 & 0 & 2 & 328 & 278 \\
24 & 0.5 & 8 & 0.02 & $1.6\times 10^6$ & 8 & 5 & 11 & 1 & 3 & 473 & 446 \\
\hline
25 & 0.5 & 20 & 0.0002 & $2\times 10^5$ & 39 & 5 & 1 & 2 & 0 & 303 & 181 \\
26 & 0.5 & 20 & 0.0002 & $4\times 10^5$ & 5 & 1 & 39 & 0 & 5 & 600 & 425 \\
27 & 0.5 & 20 & 0.0002 & $8\times 10^5$ & 33 & 18 & 5 & 1 & 4 & 877 & 733 \\
28$^\dagger$ & 0.5 & 20 & 0.0002 & $1.6\times 10^6$ & 0 & 0 & 0 & 0 & 0 & 0 & 0 \\
\hline
29 & 0.5 & 20 & 0.002 & $2\times 10^5$ & 38 & 7 & 39 & 0 & 0 & 269 & 213 \\
30 & 0.5 & 20 & 0.002 & $4\times 10^5$ & 29 & 1 & 15 & 0 & 0 & 445 & 378 \\
31 & 0.5 & 20 & 0.002 & $8\times 10^5$ & 13 & 5 & 19 & 0 & 0 & 739 & 682 \\
32 & 0.5 & 20 & 0.002 & $1.6\times 10^6$ & 0 & 0 & 0 & 0 & 0 & 1034 & 999 \\
\hline
33 & 0.5 & 20 & 0.02 & $2\times 10^5$ & 7 & 0 & 10 & 0 & 0 & 157 & 143 \\
34 & 0.5 & 20 & 0.02 & $4\times 10^5$ & 8 & 1 & 4 & 0 & 0 & 218 & 203 \\
35 & 0.5 & 20 & 0.02 & $8\times 10^5$ & 6 & 5 & 16 & 0 & 10 & 345 & 330 \\
36 & 0.5 & 20 & 0.02 & $1.6\times 10^6$ & 11 & 2 & 13 & 0 & 0 & 516 & 480 \\
\hline
37 & 1 & 2 & 0.0002 & $2\times 10^5$ & 18 & 0 & 59 & 0 & 0 & 446 & 290 \\
38 & 1 & 2 & 0.0002 & $4\times 10^5$ & 67 & 10 & 5 & 1 & 1 & 628 & 485 \\
39 & 1 & 2 & 0.0002 & $8\times 10^5$ & 0 & 0 & 0 & 0 & 0 & 907 & 785 \\
40 & 1 & 2 & 0.0002 & $1.6\times 10^6$ & 0 & 0 & 2 & 0 & 0 & 1038 & 981 \\
\hline
41 & 1 & 2 & 0.002 & $2\times 10^5$ & 28 & 0 & 14 & 0 & 1 & 285 & 231 \\
42 & 1 & 2 & 0.002 & $4\times 10^5$ & 56 & 2 & 41 & 0 & 0 & 620 & 517 \\
43 & 1 & 2 & 0.002 & $8\times 10^5$ & 0 & 0 & 4 & 0 & 0 & 704 & 629 \\
44 & 1 & 2 & 0.002 & $1.6\times 10^6$ & 0 & 0 & 0 & 0 & 0 & 807 & 753 \\
\hline
45 & 1 & 2 & 0.02 & $2\times 10^5$ & 2 & 0 & 0 & 0 & 0 & 132 & 119 \\
46 & 1 & 2 & 0.02 & $4\times 10^5$ & 7 & 1 & 38 & 0 & 5 & 230 & 216 \\
47 & 1 & 2 & 0.02 & $8\times 10^5$ & 13 & 2 & 10 & 0 & 0 & 308 & 294 \\
48 & 1 & 2 & 0.02 & $1.6\times 10^6$ & 0 & 0 & 7 & 0 & 0 & 559 & 537 \\
\hline
49 & 1 & 8 & 0.0002 & $2\times 10^5$ & 33 & 3 & 6 & 0 & 0 & 486 & 284 \\
50 & 1 & 8 & 0.0002 & $4\times 10^5$ & 38 & 4 & 24 & 0 & 1 & 738 & 582 \\
51 & 1 & 8 & 0.0002 & $8\times 10^5$ & 0 & 0 & 1 & 0 & 0 & 789 & 712 \\
52 & 1 & 8 & 0.0002 & $1.6\times 10^6$ & 0 & 0 & 0 & 0 & 0 & 978 & 906 \\
\hline
53 & 1 & 8 & 0.002 & $2\times 10^5$ & 22 & 0 & 42 & 0 & 0 & 329 & 275 \\
54 & 1 & 8 & 0.002 & $4\times 10^5$ & 1 & 0 & 2 & 0 & 0 & 571 & 491 \\
55 & 1 & 8 & 0.002 & $8\times 10^5$ & 0 & 0 & 3 & 0 & 0 & 879 & 793 \\
56 & 1 & 8 & 0.002 & $1.6\times 10^6$ & 0 & 0 & 0 & 0 & 0 & 962 & 916 \\
\hline
57 & 1 & 8 & 0.02 & $2\times 10^5$ & 7 & 0 & 1 & 0 & 0 & 210 & 196 \\
58 & 1 & 8 & 0.02 & $4\times 10^5$ & 13 & 3 & 2 & 0 & 0 & 325 & 315 \\
59 & 1 & 8 & 0.02 & $8\times 10^5$ & 3 & 1 & 24 & 0 & 0 & 411 & 394 \\
60 & 1 & 8 & 0.02 & $1.6\times 10^6$ & 0 & 0 & 7 & 0 & 0 & 486 & 471 \\
\hline
61 & 1 & 20 & 0.0002 & $2\times 10^5$ & 46 & 3 & 5 & 0 & 2 & 421 & 276 \\
62 & 1 & 20 & 0.0002 & $4\times 10^5$ & 40 & 11 & 15 & 1 & 2 & 809 & 523 \\
63 & 1 & 20 & 0.0002 & $8\times 10^5$ & 0 & 0 & 5 & 0 & 0 & 963 & 856 \\
64 & 1 & 20 & 0.0002 & $1.6\times 10^6$ & 0 & 0 & 1 & 0 & 0 & 893 & 849 \\
\hline
65 & 1 & 20 & 0.002 & $2\times 10^5$ & 14 & 0 & 26 & 0 & 0 & 415 & 352 \\
66 & 1 & 20 & 0.002 & $4\times 10^5$ & 4 & 1 & 4 & 0 & 0 & 634 & 516 \\
67 & 1 & 20 & 0.002 & $8\times 10^5$ & 0 & 0 & 1 & 0 & 0 & 745 & 671 \\
68 & 1 & 20 & 0.002 & $1.6\times 10^6$ & 0 & 0 & 0 & 0 & 0 & 1016 & 960 \\
\hline
69 & 1 & 20 & 0.02 & $2\times 10^5$ & 10 & 0 & 1 & 0 & 0 & 178 & 163 \\
70 & 1 & 20 & 0.02 & $4\times 10^5$ & 8 & 1 & 9 & 0 & 0 & 300 & 282 \\
71 & 1 & 20 & 0.02 & $8\times 10^5$ & 3 & 0 & 4 & 0 & 0 & 399 & 383 \\
72 & 1 & 20 & 0.02 & $1.6\times 10^6$ & 0 & 0 & 5 & 0 & 0 & 484 & 474 \\
\hline
73 & 2 & 2 & 0.0002 & $2\times 10^5$ & 4 & 0 & 33 & 0 & 0 & 415 & 227 \\
74 & 2 & 2 & 0.0002 & $4\times 10^5$ & 25 & 3 & 9 & 0 & 0 & 755 & 571 \\
75 & 2 & 2 & 0.0002 & $8\times 10^5$ & 0 & 0 & 0 & 0 & 0 & 908 & 734 \\
76 & 2 & 2 & 0.0002 & $1.6\times 10^6$ & 0 & 0 & 0 & 0 & 0 & 1055 & 960 \\
\hline
77 & 2 & 2 & 0.002 & $2\times 10^5$ & 3 & 0 & 5 & 0 & 0 & 472 & 333 \\
78 & 2 & 2 & 0.002 & $4\times 10^5$ & 3 & 0 & 7 & 0 & 1 & 731 & 551 \\
79 & 2 & 2 & 0.002 & $8\times 10^5$ & 0 & 0 & 0 & 0 & 0 & 786 & 650 \\
80 & 2 & 2 & 0.002 & $1.6\times 10^6$ & 0 & 0 & 1 & 0 & 0 & 855 & 765 \\
\hline
81 & 2 & 2 & 0.02 & $2\times 10^5$ & 0 & 0 & 1 & 0 & 0 & 231 & 218 \\
82 & 2 & 2 & 0.02 & $4\times 10^5$ & 2 & 0 & 2 & 0 & 0 & 245 & 238 \\
83 & 2 & 2 & 0.02 & $8\times 10^5$ & 0 & 0 & 2 & 0 & 0 & 347 & 346 \\
84 & 2 & 2 & 0.02 & $1.6\times 10^6$ & 0 & 0 & 0 & 0 & 0 & 450 & 446 \\
\hline
85 & 2 & 8 & 0.0002 & $2\times 10^5$ & 2 & 0 & 2 & 0 & 0 & 410 & 257 \\
86 & 2 & 8 & 0.0002 & $4\times 10^5$ & 0 & 0 & 0 & 0 & 0 & 688 & 488 \\
87 & 2 & 8 & 0.0002 & $8\times 10^5$ & 0 & 0 & 2 & 0 & 0 & 783 & 640 \\
88 & 2 & 8 & 0.0002 & $1.6\times 10^6$ & 0 & 0 & 0 & 0 & 0 & 779 & 715 \\
\hline
89 & 2 & 8 & 0.002 & $2\times 10^5$ & 18 & 0 & 3 & 0 & 0 & 398 & 268 \\
90 & 2 & 8 & 0.002 & $4\times 10^5$ & 0 & 0 & 3 & 0 & 0 & 748 & 600 \\
91 & 2 & 8 & 0.002 & $8\times 10^5$ & 0 & 0 & 0 & 0 & 0 & 725 & 583 \\
92 & 2 & 8 & 0.002 & $1.6\times 10^6$ & 0 & 0 & 0 & 0 & 0 & 810 & 706 \\
\hline
93 & 2 & 8 & 0.02 & $2\times 10^5$ & 0 & 0 & 0 & 0 & 0 & 160 & 145 \\
94 & 2 & 8 & 0.02 & $4\times 10^5$ & 0 & 0 & 0 & 0 & 0 & 288 & 282 \\
95 & 2 & 8 & 0.02 & $8\times 10^5$ & 0 & 0 & 0 & 0 & 0 & 367 & 359 \\
96 & 2 & 8 & 0.02 & $1.6\times 10^6$ & 0 & 0 & 0 & 0 & 0 & 437 & 434 \\
\hline
97 & 2 & 20 & 0.0002 & $2\times 10^5$ & 2 & 0 & 5 & 0 & 1 & 510 & 306 \\
98 & 2 & 20 & 0.0002 & $4\times 10^5$ & 0 & 0 & 3 & 0 & 0 & 690 & 504 \\
99 & 2 & 20 & 0.0002 & $8\times 10^5$ & 0 & 0 & 1 & 0 & 0 & 817 & 686 \\
100 & 2 & 20 & 0.0002 & $1.6\times 10^6$ & 0 & 0 & 7 & 0 & 0 & 857 & 796 \\
\hline
101 & 2 & 20 & 0.002 & $2\times 10^5$ & 0 & 0 & 4 & 0 & 0 & 337 & 222 \\
102 & 2 & 20 & 0.002 & $4\times 10^5$ & 0 & 0 & 1 & 0 & 0 & 634 & 481 \\
103 & 2 & 20 & 0.002 & $8\times 10^5$ & 0 & 0 & 2 & 0 & 0 & 755 & 616 \\
104 & 2 & 20 & 0.002 & $1.6\times 10^6$ & 0 & 0 & 0 & 0 & 0 & 931 & 822 \\
\hline
105 & 2 & 20 & 0.02 & $2\times 10^5$ & 0 & 0 & 2 & 0 & 0 & 195 & 186 \\
106 & 2 & 20 & 0.02 & $4\times 10^5$ & 0 & 0 & 0 & 0 & 0 & 263 & 257 \\
107 & 2 & 20 & 0.02 & $8\times 10^5$ & 0 & 0 & 1 & 0 & 0 & 295 & 293 \\
108 & 2 & 20 & 0.02 & $1.6\times 10^6$ & 0 & 0 & 5 & 0 & 0 & 443 & 437 \\
\hline
109 & 4 & 2 & 0.0002 & $2\times 10^5$ & 0 & 0 & 0 & 0 & 0 & 211 & 107 \\
110 & 4 & 2 & 0.0002 & $4\times 10^5$ & 0 & 0 & 9 & 0 & 0 & 399 & 268 \\
111 & 4 & 2 & 0.0002 & $8\times 10^5$ & 0 & 0 & 0 & 0 & 0 & 469 & 328 \\
112 & 4 & 2 & 0.0002 & $1.6\times 10^6$ & 0 & 0 & 0 & 0 & 0 & 797 & 710 \\
\hline
113 & 4 & 2 & 0.002 & $2\times 10^5$ & 0 & 0 & 1 & 0 & 0 & 101 & 37 \\
114 & 4 & 2 & 0.002 & $4\times 10^5$ & 0 & 0 & 2 & 0 & 0 & 358 & 241 \\
115 & 4 & 2 & 0.002 & $8\times 10^5$ & 0 & 0 & 0 & 0 & 0 & 555 & 392 \\
116 & 4 & 2 & 0.002 & $1.6\times 10^6$ & 0 & 0 & 0 & 0 & 0 & 722 & 597 \\
\hline
117 & 4 & 2 & 0.02 & $2\times 10^5$ & 0 & 0 & 0 & 0 & 0 & 103 & 100 \\
118 & 4 & 2 & 0.02 & $4\times 10^5$ & 0 & 0 & 0 & 0 & 0 & 229 & 225 \\
119 & 4 & 2 & 0.02 & $8\times 10^5$ & 0 & 0 & 0 & 0 & 0 & 288 & 283 \\
120 & 4 & 2 & 0.02 & $1.6\times 10^6$ & 0 & 0 & 0 & 0 & 0 & 333 & 330 \\
\hline
121 & 4 & 8 & 0.0002 & $2\times 10^5$ & 1 & 0 & 23 & 0 & 0 & 376 & 170 \\
122 & 4 & 8 & 0.0002 & $4\times 10^5$ & 0 & 0 & 2 & 0 & 0 & 629 & 425 \\
123 & 4 & 8 & 0.0002 & $8\times 10^5$ & 0 & 0 & 5 & 0 & 0 & 664 & 543 \\
124 & 4 & 8 & 0.0002 & $1.6\times 10^6$ & 0 & 0 & 0 & 0 & 0 & 558 & 487 \\
\hline
125 & 4 & 8 & 0.002 & $2\times 10^5$ & 0 & 0 & 4 & 0 & 0 & 381 & 225 \\
126 & 4 & 8 & 0.002 & $4\times 10^5$ & 0 & 0 & 0 & 0 & 0 & 567 & 434 \\
127 & 4 & 8 & 0.002 & $8\times 10^5$ & 0 & 0 & 0 & 0 & 0 & 817 & 622 \\
128 & 4 & 8 & 0.002 & $1.6\times 10^6$ & 0 & 0 & 1 & 0 & 0 & 876 & 728 \\
\hline
129 & 4 & 8 & 0.02 & $2\times 10^5$ & 0 & 0 & 0 & 0 & 0 & 157 & 149 \\
130 & 4 & 8 & 0.02 & $4\times 10^5$ & 0 & 0 & 0 & 0 & 0 & 226 & 223 \\
131 & 4 & 8 & 0.02 & $8\times 10^5$ & 0 & 0 & 0 & 0 & 0 & 206 & 204 \\
132 & 4 & 8 & 0.02 & $1.6\times 10^6$ & 0 & 0 & 0 & 0 & 0 & 321 & 318 \\
\hline
133 & 4 & 20 & 0.0002 & $2\times 10^5$ & 0 & 0 & 12 & 0 & 0 & 432 & 254 \\
134 & 4 & 20 & 0.0002 & $4\times 10^5$ & 0 & 0 & 7 & 0 & 0 & 613 & 411 \\
135 & 4 & 20 & 0.0002 & $8\times 10^5$ & 0 & 0 & 0 & 0 & 0 & 640 & 498 \\
136 & 4 & 20 & 0.0002 & $1.6\times 10^6$ & 0 & 0 & 0 & 0 & 0 & 648 & 549 \\
\hline
137 & 4 & 20 & 0.002 & $2\times 10^5$ & 0 & 0 & 3 & 0 & 0 & 348 & 206 \\
138 & 4 & 20 & 0.002 & $4\times 10^5$ & 0 & 0 & 5 & 0 & 0 & 574 & 409 \\
139 & 4 & 20 & 0.002 & $8\times 10^5$ & 0 & 0 & 2 & 0 & 0 & 712 & 517 \\
140 & 4 & 20 & 0.002 & $1.6\times 10^6$ & 0 & 0 & 0 & 0 & 0 & 614 & 512 \\
\hline
141 & 4 & 20 & 0.02 & $2\times 10^5$ & 0 & 0 & 1 & 0 & 0 & 146 & 139 \\
142 & 4 & 20 & 0.02 & $4\times 10^5$ & 0 & 0 & 0 & 0 & 0 & 153 & 152 \\
143 & 4 & 20 & 0.02 & $8\times 10^5$ & 0 & 0 & 0 & 0 & 0 & 230 & 227 \\
144 & 4 & 20 & 0.02 & $1.6\times 10^6$ & 0 & 0 & 1 & 0 & 0 & 338 & 337 \\
\hline
145 & 1 & 20 & 0.0002 & $3.2\times 10^6$ & 0 & 0 & 0 & 0 & 0 & 509 & 491 \\
146 & 2 & 20 & 0.0002 & $3.2\times 10^6$ & 0 & 0 & 0 & 0 & 0 & 687 & 667 \\
147 & 1 & 20 & 0.02 & $3.2\times 10^6$ & 0 & 0 & 0 & 0 & 0 & 417 & 408 \\
148 & 2 & 20 & 0.02 & $3.2\times 10^6$ & 0 & 0 & 0 & 0 & 0 & 496 & 490 \\
\hline
\enddata
\tablecomments{Models marked with a dagger ($^\dagger$) indicates the model was stopped due to onset of collisional runaway \citep[see][for details]{kremer2020}.}
\end{deluxetable*}

\bibliographystyle{aasjournal}
\bibliography{refs}

\begin{thebibliography}{}
\expandafter\ifx\csname natexlab\endcsname\relax\def\natexlab#1{#1}\fi
\providecommand{\url}[1]{\href{#1}{#1}}
\providecommand{\dodoi}[1]{doi:~\href{http://doi.org/#1}{\nolinkurl{#1}}}
\providecommand{\doeprint}[1]{\href{http://ascl.net/#1}{\nolinkurl{http://ascl.net/#1}}}
\providecommand{\doarXiv}[1]{\href{https://arxiv.org/abs/#1}{\nolinkurl{https://arxiv.org/abs/#1}}}

\bibitem[{{Aarseth} \& {Heggie}(1976)}]{Aarseth1976}
{Aarseth}, S.~J., \& {Heggie}, D.~C. 1976, \aap, 53, 259

\bibitem[{{Anderson} {et~al.}(2017){Anderson}, {Lai}, \& {Storch}}]{anders2017}
{Anderson}, K.~R., {Lai}, D., \& {Storch}, N.~I. 2017, \mnras, 467, 3066,
  \dodoi{10.1093/mnras/stx293}

\bibitem[{{Anderson} {et~al.}(2016){Anderson}, {Storch}, \& {Lai}}]{anders2016}
{Anderson}, K.~R., {Storch}, N.~I., \& {Lai}, D. 2016, \mnras, 456, 3671,
  \dodoi{10.1093/mnras/stv2906}

\bibitem[{{Antognini} \& {Thompson}(2016)}]{antogn2016}
{Antognini}, J. M.~O., \& {Thompson}, T.~A. 2016, \mnras, 456, 4219,
  \dodoi{10.1093/mnras/stv2938}

\bibitem[{{Antonini} {et~al.}(2016){Antonini}, {Chatterjee}, {Rodriguez},
  {Morscher}, {Pattabiraman}, {Kalogera}, \& {Rasio}}]{antoninietal2016}
{Antonini}, F., {Chatterjee}, S., {Rodriguez}, C.~L., {et~al.} 2016, \apj, 816,
  65, \dodoi{10.3847/0004-637X/816/2/65}

\bibitem[{{Arzoumanian} {et~al.}(1996){Arzoumanian}, {Joshi}, {Rasio}, \&
  {Thorsett}}]{arz1996}
{Arzoumanian}, Z., {Joshi}, K., {Rasio}, F.~A., \& {Thorsett}, S.~E. 1996, in
  Astronomical Society of the Pacific Conference Series, Vol. 105, IAU Colloq.
  160: Pulsars: Problems and Progress, ed. S.~{Johnston}, M.~A. {Walker}, \&
  M.~{Bailes}, 525--530

\bibitem[{{Askar} {et~al.}(2017){Askar}, {Szkudlarek}, {Gondek-Rosi{\'n}ska},
  {Giersz}, \& {Bulik}}]{Askar2017}
{Askar}, A., {Szkudlarek}, M., {Gondek-Rosi{\'n}ska}, D., {Giersz}, M., \&
  {Bulik}, T. 2017, \mnras, 464, L36, \dodoi{10.1093/mnrasl/slw177}

\bibitem[{{Banerjee}(2017)}]{Banerjee2017}
{Banerjee}, S. 2017, \mnras, 467, 524, \dodoi{10.1093/mnras/stw3392}

\bibitem[{Banerjee {et~al.}(2010)Banerjee, Baumgardt, \& Kroupa}]{Banerjee2010}
Banerjee, S., Baumgardt, H., \& Kroupa, P. 2010, Mon.~Not.~R.~Astron.~Soc, 402,
  371, \dodoi{10.1111/j.1365-2966.2009.15880.x}

\bibitem[{{Baumgardt} \& {Hilker}(2018)}]{Baumgardt2018}
{Baumgardt}, H., \& {Hilker}, M. 2018, \mnras, 478, 1520,
  \dodoi{10.1093/mnras/sty1057}

\bibitem[{{Belczynski} {et~al.}(2002){Belczynski}, {Kalogera}, \&
  {Bulik}}]{Belczynski2002}
{Belczynski}, K., {Kalogera}, V., \& {Bulik}, T. 2002, \apj, 572, 407,
  \dodoi{10.1086/340304}

\bibitem[{Belczynski {et~al.}(2016)Belczynski, Heger, Gladysz, Ruiter, Woosley,
  Wiktorowicz, Chen, Bulik, O’Shaughnessy, Holz, Fryer, \&
  Berti}]{Belczynski2016b}
Belczynski, K., Heger, A., Gladysz, W., {et~al.} 2016, Astronomy {\&}
  Astrophysics, 594, A97, \dodoi{10.1051/0004-6361/201628980}

\bibitem[{{Binney} \& {Tremaine}(2008)}]{binneytremaine2008}
{Binney}, J., \& {Tremaine}, S. 2008, {Galactic Dynamics: Second Edition}

\bibitem[{{Bouvier} {et~al.}(2001){Bouvier}, {Duch{\^e}ne}, {Mermilliod}, \&
  {Simon}}]{bouvier01}
{Bouvier}, J., {Duch{\^e}ne}, G., {Mermilliod}, J.-C., \& {Simon}, T. 2001,
  \aap, 375, 989, \dodoi{10.1051/0004-6361:20010915}

\bibitem[{{Bouvier} {et~al.}(1997){Bouvier}, {Rigaut}, \& {Nadeau}}]{bouvier97}
{Bouvier}, J., {Rigaut}, F., \& {Nadeau}, D. 1997, \aap, 323, 139

\bibitem[{Chatterjee {et~al.}(2010)Chatterjee, Fregeau, Umbreit, \&
  Rasio}]{Chatterjee2010}
Chatterjee, S., Fregeau, J.~M., Umbreit, S., \& Rasio, F.~A. 2010, The
  Astrophysical Journal, 719, 915

\bibitem[{{Chatterjee} {et~al.}(2017{\natexlab{a}}){Chatterjee}, {Rodriguez},
  {Kalogera}, \& {Rasio}}]{Chatterjee2017b}
{Chatterjee}, S., {Rodriguez}, C.~L., {Kalogera}, V., \& {Rasio}, F.~A.
  2017{\natexlab{a}}, \apjl, 836, L26, \dodoi{10.3847/2041-8213/aa5caa}

\bibitem[{{Chatterjee} {et~al.}(2017{\natexlab{b}}){Chatterjee}, {Rodriguez},
  \& {Rasio}}]{Chatterjee2017a}
{Chatterjee}, S., {Rodriguez}, C.~L., \& {Rasio}, F.~A. 2017{\natexlab{b}},
  \apj, 834, 68, \dodoi{10.3847/1538-4357/834/1/68}

\bibitem[{Chatterjee {et~al.}(2013)Chatterjee, Umbreit, Fregeau, \&
  Rasio}]{Chatterjee2013}
Chatterjee, S., Umbreit, S., Fregeau, J.~M., \& Rasio, F.~A. 2013, Monthly
  Notices of the Royal Astronomical Society, 429, 2881,
  \dodoi{10.1093/mnras/sts464}

\bibitem[{{Chen} {et~al.}(2009){Chen}, {Madau}, {Sesana}, \& {Liu}}]{chen2009}
{Chen}, X., {Madau}, P., {Sesana}, A., \& {Liu}, F.~K. 2009, \apjl, 697, L149,
  \dodoi{10.1088/0004-637X/697/2/L149}

\bibitem[{{Chini} {et~al.}(2012){Chini}, {Hoffmeister}, {Nasseri}, {Stahl}, \&
  {Zinnecker}}]{chi12}
{Chini}, R., {Hoffmeister}, V.~H., {Nasseri}, A., {Stahl}, O., \& {Zinnecker},
  H. 2012, \mnras, 424, 1925, \dodoi{10.1111/j.1365-2966.2012.21317.x}

\bibitem[{Clark(1975)}]{Clark1975}
Clark, G. 1975, \apj, 199, L143

\bibitem[{{Dehnen} \& {Binney}(1998)}]{Dehnen1998}
{Dehnen}, W., \& {Binney}, J. 1998, \mnras, 294, 429,
  \dodoi{10.1046/j.1365-8711.1998.01282.x}

\bibitem[{{Duch{\^e}ne} \& {Kraus}(2013)}]{duc13}
{Duch{\^e}ne}, G., \& {Kraus}, A. 2013, \araa, 51, 269,
  \dodoi{10.1146/annurev-astro-081710-102602}

\bibitem[{{Duquennoy} \& {Mayor}(1991)}]{DuquennoyMayor1991}
{Duquennoy}, A., \& {Mayor}, M. 1991, \aap, 500, 337

\bibitem[{{Eggleton}(1983)}]{egg83}
{Eggleton}, P.~P. 1983, \apj, 268, 368, \dodoi{10.1086/160960}

\bibitem[{{Eggleton} \& {Kiseleva-Eggleton}(2001)}]{egg2001}
{Eggleton}, P.~P., \& {Kiseleva-Eggleton}, L. 2001, \apj, 562, 1012,
  \dodoi{10.1086/323843}

\bibitem[{{El-Badry} {et~al.}(2018){El-Badry}, {Quataert}, {Weisz}, {Choksi},
  \& {Boylan-Kolchin}}]{El-Badry2018}
{El-Badry}, K., {Quataert}, E., {Weisz}, D.~R., {Choksi}, N., \&
  {Boylan-Kolchin}, M. 2018, ArXiv e-prints.
\newblock \doarXiv{1805.03652}

\bibitem[{{Fabrycky} \& {Tremaine}(2007)}]{fabt2007}
{Fabrycky}, D., \& {Tremaine}, S. 2007, \apj, 669, 1298, \dodoi{10.1086/521702}

\bibitem[{{Ford} {et~al.}(2000){Ford}, {Kozinsky}, \& {Rasio}}]{ford2000}
{Ford}, E.~B., {Kozinsky}, B., \& {Rasio}, F.~A. 2000, \apj, 535, 385,
  \dodoi{10.1086/308815}

\bibitem[{{Fragione} {et~al.}(2019{\natexlab{a}}){Fragione}, {Grishin},
  {Leigh}, {Perets}, \& {Perna}}]{fragrish2019}
{Fragione}, G., {Grishin}, E., {Leigh}, N. W.~C., {Perets}, H.~B., \& {Perna},
  R. 2019{\natexlab{a}}, \mnras, 488, 47, \dodoi{10.1093/mnras/stz1651}

\bibitem[{{Fragione} \& {Kocsis}(2018)}]{Fragione2018b}
{Fragione}, G., \& {Kocsis}, B. 2018, Physical Review Letters, 121, 161103,
  \dodoi{10.1103/PhysRevLett.121.161103}

\bibitem[{{Fragione} \& {Kocsis}(2019)}]{Fragkoc2019}
---. 2019, \mnras, 486, 4781, \dodoi{10.1093/mnras/stz1175}

\bibitem[{{Fragione} \& {Kocsis}(2020)}]{fragk2020}
---. 2020, \mnras, 493, 3920, \dodoi{10.1093/mnras/staa443}

\bibitem[{{Fragione} \& {Leigh}(2018)}]{fragleigh2018}
{Fragione}, G., \& {Leigh}, N. 2018, \mnras, 479, 3181,
  \dodoi{10.1093/mnras/sty1600}

\bibitem[{{Fragione} {et~al.}(2019{\natexlab{b}}){Fragione}, {Leigh}, \&
  {Perna}}]{Fragleipern2019}
{Fragione}, G., {Leigh}, N. W.~C., \& {Perna}, R. 2019{\natexlab{b}}, \mnras,
  488, 2825, \dodoi{10.1093/mnras/stz1803}

\bibitem[{{Fragione} {et~al.}(2019{\natexlab{c}}){Fragione}, {Leigh}, {Perna},
  \& {Kocsis}}]{fragetal2019}
{Fragione}, G., {Leigh}, N. W.~C., {Perna}, R., \& {Kocsis}, B.
  2019{\natexlab{c}}, \mnras, 489, 727, \dodoi{10.1093/mnras/stz2213}

\bibitem[{{Fragione} {et~al.}(2019{\natexlab{d}}){Fragione}, {Loeb}, \&
  {Ginsburg}}]{frag2019}
{Fragione}, G., {Loeb}, A., \& {Ginsburg}, I. 2019{\natexlab{d}}, \mnras, 483,
  648, \dodoi{10.1093/mnras/sty3194}

\bibitem[{{Fragione} {et~al.}(2019{\natexlab{e}}){Fragione}, {Metzger},
  {Perna}, {Leigh}, \& {Kocsis}}]{fragmet2019}
{Fragione}, G., {Metzger}, B.~D., {Perna}, R., {Leigh}, N. W.~C., \& {Kocsis},
  B. 2019{\natexlab{e}}, arXiv e-prints, arXiv:1908.00987.
\newblock \doarXiv{1908.00987}

\bibitem[{{Fragione} {et~al.}(2018){Fragione}, {Pavl{\'{\i}}k}, \&
  {Banerjee}}]{Fragione2018a}
{Fragione}, G., {Pavl{\'{\i}}k}, V., \& {Banerjee}, S. 2018, \mnras, 480, 4955,
  \dodoi{10.1093/mnras/sty2234}

\bibitem[{{Fregeau} {et~al.}(2004){Fregeau}, {Cheung}, {Portegies Zwart}, \&
  {Rasio}}]{Fregeau2004}
{Fregeau}, J.~M., {Cheung}, P., {Portegies Zwart}, S.~F., \& {Rasio}, F.~A.
  2004, \mnras, 352, 1, \dodoi{10.1111/j.1365-2966.2004.07914.x}

\bibitem[{Fregeau {et~al.}(2003)Fregeau, Gurkan, Joshi, \& Rasio}]{Fregeau2003}
Fregeau, J.~M., Gurkan, M.~A., Joshi, K.~J., \& Rasio, F.~A. 2003, arXiv,
  astro-ph, 772

\bibitem[{{Fregeau} \& {Rasio}(2007)}]{Fregeau2007}
{Fregeau}, J.~M., \& {Rasio}, F.~A. 2007, \apj, 658, 1047,
  \dodoi{10.1086/511809}

\bibitem[{{Fryer} {et~al.}(2012){Fryer}, {Belczynski}, {Wiktorowicz},
  {Dominik}, {Kalogera}, \& {Holz}}]{Fryer2012}
{Fryer}, C.~L., {Belczynski}, K., {Wiktorowicz}, G., {et~al.} 2012, \apj, 749,
  91, \dodoi{10.1088/0004-637X/749/1/91}

\bibitem[{Fryer \& Kalogera(2001)}]{Fryer2001}
Fryer, C.~L., \& Kalogera, V. 2001, The Astrophysical Journal, 554, 548,
  \dodoi{10.1086/321359}

\bibitem[{{Fryer} {et~al.}(1999){Fryer}, {Woosley}, {Herant}, \&
  {Davies}}]{fryer1999}
{Fryer}, C.~L., {Woosley}, S.~E., {Herant}, M., \& {Davies}, M.~B. 1999, \apj,
  520, 650, \dodoi{10.1086/307467}

\bibitem[{{Giesler} {et~al.}(2018){Giesler}, {Clausen}, \& {Ott}}]{Giesler2018}
{Giesler}, M., {Clausen}, D., \& {Ott}, C.~D. 2018, \mnras, 477, 1853,
  \dodoi{10.1093/mnras/sty659}

\bibitem[{{Grindlay} {et~al.}(1988){Grindlay}, {Bailyn}, {Cohn}, {Lugger},
  {Thorstensen}, \& {Wegner}}]{grindlay88}
{Grindlay}, J.~E., {Bailyn}, C.~D., {Cohn}, H., {et~al.} 1988, \apjl, 334, L25,
  \dodoi{10.1086/185305}

\bibitem[{{Grishin} {et~al.}(2018){Grishin}, {Perets}, \&
  {Fragione}}]{grish2018}
{Grishin}, E., {Perets}, H.~B., \& {Fragione}, G. 2018, \mnras, 481, 4907,
  \dodoi{10.1093/mnras/sty2477}

\bibitem[{{Hamers} {et~al.}(2018){Hamers}, {Bar-Or}, {Petrovich}, \&
  {Antonini}}]{hamers2018}
{Hamers}, A.~S., {Bar-Or}, B., {Petrovich}, C., \& {Antonini}, F. 2018, \apj,
  865, 2, \dodoi{10.3847/1538-4357/aadae2}

\bibitem[{{Harris}(1996)}]{Harris1996}
{Harris}, W.~E. 1996, \aj, 112, 1487, \dodoi{10.1086/118116}

\bibitem[{{Heggie} \& {Hut}(2003)}]{HeggieHut2003}
{Heggie}, D., \& {Hut}, P. 2003, {The Gravitational Million-Body Problem: A
  Multidisciplinary Approach to Star Cluster Dynamics}

\bibitem[{Heggie(1975)}]{Heggie1975}
Heggie, D.~C. 1975, Mon.~Not.~R.~Astron.~Soc, 173, 729

\bibitem[{{Heinke} {et~al.}(2005){Heinke}, {Grindlay}, {Edmonds}, {Cohn},
  {Lugger}, {Camilo}, {Bogdanov}, \& {Freire}}]{Heinke2005}
{Heinke}, C.~O., {Grindlay}, J.~E., {Edmonds}, P.~D., {et~al.} 2005, \apj, 625,
  796, \dodoi{10.1086/429899}

\bibitem[{H{\'{e}}non(1971{\natexlab{a}})}]{Henon1971a}
H{\'{e}}non, M. 1971{\natexlab{a}}, Astrophysics and Space Science, 13, 284

\bibitem[{H{\'{e}}non(1971{\natexlab{b}})}]{Henon1971b}
---. 1971{\natexlab{b}}, Astrophysics and Space Science, 14, 151,
  \dodoi{10.1007/BF00649201}

\bibitem[{{Hoang} {et~al.}(2018){Hoang}, {Naoz}, {Kocsis}, {Rasio}, \&
  {Dosopoulou}}]{hoang2018}
{Hoang}, B.-M., {Naoz}, S., {Kocsis}, B., {Rasio}, F.~A., \& {Dosopoulou}, F.
  2018, \apj, 856, 140, \dodoi{10.3847/1538-4357/aaafce}

\bibitem[{Hobbs {et~al.}(2005)Hobbs, Lorimer, Lyne, \& Kramer}]{Hobbs2005}
Hobbs, G., Lorimer, D.~R., Lyne, A.~G., \& Kramer, M. 2005, Monthly Notices of
  the Royal Astronomical Society, 360, 974,
  \dodoi{10.1111/j.1365-2966.2005.09087.x}

\bibitem[{{Hong} {et~al.}(2018){Hong}, {Vesperini}, {Askar}, {Giersz},
  {Szkudlarek}, \& {Bulik}}]{Hong2018}
{Hong}, J., {Vesperini}, E., {Askar}, A., {et~al.} 2018, \mnras, 480, 5645,
  \dodoi{10.1093/mnras/sty2211}

\bibitem[{{Hurley} {et~al.}(2000){Hurley}, {Pols}, \& {Tout}}]{Hurley2000}
{Hurley}, J.~R., {Pols}, O.~R., \& {Tout}, C.~A. 2000, \mnras, 315, 543,
  \dodoi{10.1046/j.1365-8711.2000.03426.x}

\bibitem[{{Hurley} {et~al.}(2002){Hurley}, {Tout}, \& {Pols}}]{Hurley2002}
{Hurley}, J.~R., {Tout}, C.~A., \& {Pols}, O.~R. 2002, \mnras, 329, 897,
  \dodoi{10.1046/j.1365-8711.2002.05038.x}

\bibitem[{{Hut}(1981)}]{hut1981}
{Hut}, P. 1981, \aap, 99, 126

\bibitem[{{Ivanova}(2013)}]{Ivanova2013}
{Ivanova}, N. 2013, \memsai, 84, 123.
\newblock \doarXiv{1301.2203}

\bibitem[{{Ivanova} {et~al.}(2010){Ivanova}, {Chaichenets}, {Fregeau},
  {Heinke}, {Lombardi}, \& {Woods}}]{Ivanova2010}
{Ivanova}, N., {Chaichenets}, S., {Fregeau}, J., {et~al.} 2010, \apj, 717, 948,
  \dodoi{10.1088/0004-637X/717/2/948}

\bibitem[{{Ivanova} {et~al.}(2008){Ivanova}, {Heinke}, {Rasio}, {Belczynski},
  \& {Fregeau}}]{Ivanova2008}
{Ivanova}, N., {Heinke}, C.~O., {Rasio}, F.~A., {Belczynski}, K., \& {Fregeau},
  J.~M. 2008, \mnras, 386, 553, \dodoi{10.1111/j.1365-2966.2008.13064.x}

\bibitem[{Joshi {et~al.}(2001)Joshi, Nave, \& Rasio}]{Joshi2001}
Joshi, K.~J., Nave, C.~P., \& Rasio, F.~A. 2001, The Astrophysical Journal,
  550, 691

\bibitem[{Joshi {et~al.}(2000)Joshi, Rasio, Zwart, \&
  Portegies~Zwart}]{Joshi2000}
Joshi, K.~J., Rasio, F.~A., Zwart, S.~P., \& Portegies~Zwart, S. 2000, The
  Astrophysical Journal, 540, 969, \dodoi{10.1086/309350}

\bibitem[{{King}(1962)}]{King1962}
{King}, I. 1962, \aj, 67, 471, \dodoi{10.1086/108756}

\bibitem[{{Kiseleva} {et~al.}(1998){Kiseleva}, {Eggleton}, \&
  {Mikkola}}]{kisel1998}
{Kiseleva}, L.~G., {Eggleton}, P.~P., \& {Mikkola}, S. 1998, \mnras, 300, 292,
  \dodoi{10.1046/j.1365-8711.1998.01903.x}

\bibitem[{{Kozai}(1962)}]{koz62}
{Kozai}, Y. 1962, \aj, 67, 591, \dodoi{10.1086/108790}

\bibitem[{{Kraus} {et~al.}(2011){Kraus}, {Ireland}, {Martinache}, \&
  {Hillenbrand}}]{kraus11}
{Kraus}, A.~L., {Ireland}, M.~J., {Martinache}, F., \& {Hillenbrand}, L.~A.
  2011, \apj, 731, 8, \dodoi{10.1088/0004-637X/731/1/8}

\bibitem[{{Kremer} {et~al.}(2018){Kremer}, {Chatterjee}, {Rodriguez}, \&
  {Rasio}}]{Kremer2018a}
{Kremer}, K., {Chatterjee}, S., {Rodriguez}, C.~L., \& {Rasio}, F.~A. 2018,
  \apj, 852, 29, \dodoi{10.3847/1538-4357/aa99df}

\bibitem[{{Kremer} {et~al.}(2019{\natexlab{a}}){Kremer}, {Chatterjee}, {Ye},
  {Rodriguez}, \& {Rasio}}]{Kremer2019a}
{Kremer}, K., {Chatterjee}, S., {Ye}, C.~S., {Rodriguez}, C.~L., \& {Rasio},
  F.~A. 2019{\natexlab{a}}, \apj, 871, 38, \dodoi{10.3847/1538-4357/aaf646}

\bibitem[{{Kremer} {et~al.}(2019{\natexlab{b}}){Kremer}, {D'Orazio}, {Samsing},
  {Chatterjee}, \& {Rasio}}]{kremdoraz2019}
{Kremer}, K., {D'Orazio}, D.~J., {Samsing}, J., {Chatterjee}, S., \& {Rasio},
  F.~A. 2019{\natexlab{b}}, \apj, 885, 2, \dodoi{10.3847/1538-4357/ab44d1}

\bibitem[{{Kremer} {et~al.}(2019{\natexlab{c}}){Kremer}, {Lu}, {Rodriguez},
  {Lachat}, \& {Rasio}}]{Kremer2019c}
{Kremer}, K., {Lu}, W., {Rodriguez}, C.~L., {Lachat}, M., \& {Rasio}, F.
  2019{\natexlab{c}}, arXiv e-prints.
\newblock \doarXiv{1904.06353}

\bibitem[{{Kremer} {et~al.}(2019{\natexlab{d}}){Kremer}, {Ye}, {Chatterjee},
  {Rodriguez}, \& {Rasio}}]{Kremer2019d}
{Kremer}, K., {Ye}, C.~S., {Chatterjee}, S., {Rodriguez}, C.~L., \& {Rasio},
  F.~A. 2019{\natexlab{d}}, arXiv e-prints, arXiv:1907.12564.
\newblock \doarXiv{1907.12564}

\bibitem[{{Kremer} {et~al.}(2019{\natexlab{e}}){Kremer}, {Rodriguez},
  {Amaro-Seoane}, {Breivik}, {Chatterjee}, {Katz}, {Larson}, {Rasio},
  {Samsing}, {Ye}, \& {Zevin}}]{Kremer2019b}
{Kremer}, K., {Rodriguez}, C.~L., {Amaro-Seoane}, P., {et~al.}
  2019{\natexlab{e}}, \prd, 99, 063003, \dodoi{10.1103/PhysRevD.99.063003}

\bibitem[{{Kremer} {et~al.}(2020){Kremer}, {Ye}, {Rui}, {Weatherford},
  {Chatterjee}, {Fragione}, {Rodriguez}, {Spera}, \& {Rasio}}]{kremer2020}
{Kremer}, K., {Ye}, C.~S., {Rui}, N.~Z., {et~al.} 2020, \apjs, 247, 48,
  \dodoi{10.3847/1538-4365/ab7919}

\bibitem[{{Kroupa}(2001)}]{Kroupa2001}
{Kroupa}, P. 2001, \mnras, 322, 231, \dodoi{10.1046/j.1365-8711.2001.04022.x}

\bibitem[{{Leigh} \& {Geller}(2013)}]{leighgeller2013}
{Leigh}, N. W.~C., \& {Geller}, A.~M. 2013, \mnras, 432, 2474,
  \dodoi{10.1093/mnras/stt617}

\bibitem[{{Leigh} {et~al.}(2016){Leigh}, {Stone}, {Geller}, {Shara}, {Muddu},
  {Solano-Oropeza}, \& {Thomas}}]{leigh2016}
{Leigh}, N. W.~C., {Stone}, N.~C., {Geller}, A.~M., {et~al.} 2016, \mnras, 463,
  3311, \dodoi{10.1093/mnras/stw2178}

\bibitem[{{Leigh} {et~al.}(2020){Leigh}, {Toonen}, {Portegies Zwart}, \&
  {Perna}}]{leigh2020}
{Leigh}, N. W.~C., {Toonen}, S., {Portegies Zwart}, S.~F., \& {Perna}, R. 2020,
  \mnras, 496, 1819, \dodoi{10.1093/mnras/staa1670}

\bibitem[{{Lidov}(1962)}]{lid62}
{Lidov}, M.~L. 1962, \planss, 9, 719, \dodoi{10.1016/0032-0633(62)90129-0}

\bibitem[{{Liu} {et~al.}(2019){Liu}, {Lai}, \& {Wang}}]{liu2019}
{Liu}, B., {Lai}, D., \& {Wang}, Y.-H. 2019, \apj, 881, 41,
  \dodoi{10.3847/1538-4357/ab2dfb}

\bibitem[{{Liu} {et~al.}(2015){Liu}, {Mu{\~n}oz}, \& {Lai}}]{liu2015}
{Liu}, B., {Mu{\~n}oz}, D.~J., \& {Lai}, D. 2015, \mnras, 447, 747,
  \dodoi{10.1093/mnras/stu2396}

\bibitem[{Lyne {et~al.}(1987)Lyne, Brinklow, Middleditch, Kulkarni, Backer, \&
  Clifton}]{Lyne1987}
Lyne, A., Brinklow, A., Middleditch, J., {et~al.} 1987, \nat, 328, 399

\bibitem[{{Mardling} \& {Aarseth}(2001)}]{mard01}
{Mardling}, R.~A., \& {Aarseth}, S.~J. 2001, \mnras, 321, 398,
  \dodoi{10.1046/j.1365-8711.2001.03974.x}

\bibitem[{{Mermilliod} \& {Mayor}(1999)}]{mermilliod99}
{Mermilliod}, J.-C., \& {Mayor}, M. 1999, \aap, 352, 479

\bibitem[{{Mermilliod} {et~al.}(1992){Mermilliod}, {Rosvick}, {Duquennoy}, \&
  {Mayor}}]{mermilliod92}
{Mermilliod}, J.-C., {Rosvick}, J.~M., {Duquennoy}, A., \& {Mayor}, M. 1992,
  \aap, 265, 513

\bibitem[{{Moody} \& {Sigurdsson}(2009)}]{Moody2009}
{Moody}, K., \& {Sigurdsson}, S. 2009, \apj, 690, 1370,
  \dodoi{10.1088/0004-637X/690/2/1370}

\bibitem[{Morscher {et~al.}(2015)Morscher, Pattabiraman, Rodriguez, Rasio, \&
  Umbreit}]{Morscher2015}
Morscher, M., Pattabiraman, B., Rodriguez, C., Rasio, F.~A., \& Umbreit, S.
  2015, The Astrophysical Journal, 800, 9, \dodoi{10.1088/0004-637X/800/1/9}

\bibitem[{{Naoz}(2016)}]{naoz2016}
{Naoz}, S. 2016, \araa, 54, 441, \dodoi{10.1146/annurev-astro-081915-023315}

\bibitem[{{Naoz} \& {Fabrycky}(2014)}]{naozf2014}
{Naoz}, S., \& {Fabrycky}, D.~C. 2014, \apj, 793, 137,
  \dodoi{10.1088/0004-637X/793/2/137}

\bibitem[{{Naoz} {et~al.}(2013){Naoz}, {Farr}, {Lithwick}, {Rasio}, \&
  {Teyssandier}}]{naoz13a}
{Naoz}, S., {Farr}, W.~M., {Lithwick}, Y., {Rasio}, F.~A., \& {Teyssandier}, J.
  2013, \mnras, 431, 2155, \dodoi{10.1093/mnras/stt302}

\bibitem[{{Naoz} {et~al.}(2016){Naoz}, {Fragos}, {Geller}, {Stephan}, \&
  {Rasio}}]{naozf2016}
{Naoz}, S., {Fragos}, T., {Geller}, A., {Stephan}, A.~P., \& {Rasio}, F.~A.
  2016, \apjl, 822, L24, \dodoi{10.3847/2041-8205/822/2/L24}

\bibitem[{{Patience} {et~al.}(1998){Patience}, {Ghez}, {Reid}, {Weinberger}, \&
  {Matthews}}]{patience98}
{Patience}, J., {Ghez}, A.~M., {Reid}, I.~N., {Weinberger}, A.~J., \&
  {Matthews}, K. 1998, \aj, 115, 1972, \dodoi{10.1086/300321}

\bibitem[{Pattabiraman {et~al.}(2013)Pattabiraman, Umbreit, Liao, Choudhary,
  Kalogera, Memik, \& Rasio}]{Pattabiraman2013}
Pattabiraman, B., Umbreit, S., Liao, W.-k., {et~al.} 2013, The Astrophysical
  Journal Supplement Series, 204, 15, \dodoi{10.1088/0067-0049/204/2/15}

\bibitem[{{Perets} \& {Fabrycky}(2009)}]{peretsfab2009}
{Perets}, H.~B., \& {Fabrycky}, D.~C. 2009, \apj, 697, 1048,
  \dodoi{10.1088/0004-637X/697/2/1048}

\bibitem[{{Perets} {et~al.}(2016){Perets}, {Li}, {Lombardi}, \&
  {Milcarek}}]{perets2016}
{Perets}, H.~B., {Li}, Z., {Lombardi}, James~C., J., \& {Milcarek}, Stephen~R.,
  J. 2016, \apj, 823, 113, \dodoi{10.3847/0004-637X/823/2/113}

\bibitem[{Peters(1964)}]{Peters1964}
Peters, P. 1964, Physical Review, 136, B1224, \dodoi{10.1103/PhysRev.136.B1224}

\bibitem[{{Petrovich} \& {Antonini}(2017)}]{petrov2017}
{Petrovich}, C., \& {Antonini}, F. 2017, \apj, 846, 146,
  \dodoi{10.3847/1538-4357/aa8628}

\bibitem[{{Prodan} \& {Murray}(2012)}]{prodan12}
{Prodan}, S., \& {Murray}, N. 2012, \apj, 747, 4,
  \dodoi{10.1088/0004-637X/747/1/4}

\bibitem[{{Ransom}(2008)}]{Ransom2008}
{Ransom}, S.~M. 2008, in IAU Symposium, Vol. 246, Dynamical Evolution of Dense
  Stellar Systems, ed. E.~{Vesperini}, M.~{Giersz}, \& A.~{Sills}, 291--300

\bibitem[{{Rasio} {et~al.}(1995){Rasio}, {McMillan}, \& {Hut}}]{rasio1995}
{Rasio}, F.~A., {McMillan}, S., \& {Hut}, P. 1995, \apjl, 438, L33,
  \dodoi{10.1086/187708}

\bibitem[{{Riddle} {et~al.}(2015){Riddle}, {Tokovinin}, {Mason}, {Hartkopf},
  {Roberts}, {Baranec}, {Law}, {Bui}, {Burse}, {Das}, {Dekany}, {Kulkarni},
  {Punnadi}, {Ramaprakash}, \& {Tendulkar}}]{rid15}
{Riddle}, R.~L., {Tokovinin}, A., {Mason}, B.~D., {et~al.} 2015, \apj, 799, 4,
  \dodoi{10.1088/0004-637X/799/1/4}

\bibitem[{{Rivinius} {et~al.}(2020){Rivinius}, {Baade}, {Hadrava}, {Heida}, \&
  {Klement}}]{rivinius2020}
{Rivinius}, T., {Baade}, D., {Hadrava}, P., {Heida}, M., \& {Klement}, R. 2020,
  A\& A, 637, L3, \dodoi{10.1051/0004-6361/202038020}

\bibitem[{{Robson} {et~al.}(2019){Robson}, {Cornish}, \& {Liu}}]{Robson+2019}
{Robson}, T., {Cornish}, N.~J., \& {Liu}, C. 2019, Classical and Quantum
  Gravity, 36, 105011, \dodoi{10.1088/1361-6382/ab1101}

\bibitem[{{Rodriguez} {et~al.}(2018{\natexlab{a}}){Rodriguez}, {Amaro-Seoane},
  {Chatterjee}, {Kremer}, {Rasio}, {Samsing}, {Ye}, \&
  {Zevin}}]{Rodriguez2018b}
{Rodriguez}, C.~L., {Amaro-Seoane}, P., {Chatterjee}, S., {et~al.}
  2018{\natexlab{a}}, \prd, 98, 123005, \dodoi{10.1103/PhysRevD.98.123005}

\bibitem[{{Rodriguez} {et~al.}(2018{\natexlab{b}}){Rodriguez}, {Amaro-Seoane},
  {Chatterjee}, \& {Rasio}}]{Rodriguez2018a}
{Rodriguez}, C.~L., {Amaro-Seoane}, P., {Chatterjee}, S., \& {Rasio}, F.~A.
  2018{\natexlab{b}}, Physical Review Letters, 120, 151101,
  \dodoi{10.1103/PhysRevLett.120.151101}

\bibitem[{Rodriguez {et~al.}(2016)Rodriguez, Chatterjee, \&
  Rasio}]{Rodriguez2016a}
Rodriguez, C.~L., Chatterjee, S., \& Rasio, F.~A. 2016, Physical Review D, 93,
  084029, \dodoi{10.1103/PhysRevD.93.084029}

\bibitem[{{Rodriguez} \& {Loeb}(2018)}]{RodriguezLoeb2018}
{Rodriguez}, C.~L., \& {Loeb}, A. 2018, ArXiv e-prints.
\newblock \doarXiv{1809.01152}

\bibitem[{Rodriguez {et~al.}(2015)Rodriguez, Morscher, Pattabiraman,
  Chatterjee, Haster, \& Rasio}]{Rodriguez2015a}
Rodriguez, C.~L., Morscher, M., Pattabiraman, B., {et~al.} 2015, Physical
  Review Letters, 115, 051101, \dodoi{10.1103/PhysRevLett.115.051101}

\bibitem[{{Rodriguez} {et~al.}(2016){Rodriguez}, {Morscher}, {Wang},
  {Chatterjee}, {Rasio}, \& {Spurzem}}]{rodcomp2016}
{Rodriguez}, C.~L., {Morscher}, M., {Wang}, L., {et~al.} 2016, \mnras, 463,
  2109, \dodoi{10.1093/mnras/stw2121}

\bibitem[{{Rose} {et~al.}(2019){Rose}, {Naoz}, \& {Geller}}]{rose2019}
{Rose}, S.~C., {Naoz}, S., \& {Geller}, A.~M. 2019, \mnras, 488, 2480,
  \dodoi{10.1093/mnras/stz1846}

\bibitem[{{Samsing} \& {D'Orazio}(2018)}]{Samsing2018a}
{Samsing}, J., \& {D'Orazio}, D.~J. 2018, \mnras, \dodoi{10.1093/mnras/sty2334}

\bibitem[{{Samsing} {et~al.}(2019){Samsing}, {D'Orazio}, {Kremer}, {Rodriguez},
  \& {Askar}}]{Samsing2019_singlesingle}
{Samsing}, J., {D'Orazio}, D.~J., {Kremer}, K., {Rodriguez}, C.~L., \& {Askar},
  A. 2019, arXiv e-prints, arXiv:1907.11231.
\newblock \doarXiv{1907.11231}

\bibitem[{{Sana} {et~al.}(2014){Sana}, {Le Bouquin}, {Lacour}, {Berger},
  {Duvert}, {Gauchet}, {Norris}, {Olofsson}, {Pickel}, {Zins}, {Absil}, {de
  Koter}, {Kratter}, {Schnurr}, \& {Zinnecker}}]{san14}
{Sana}, H., {Le Bouquin}, J.-B., {Lacour}, S., {et~al.} 2014, \apjs, 215, 15,
  \dodoi{10.1088/0067-0049/215/1/15}

\bibitem[{{Sigurdsson} \& {Phinney}(1993)}]{sigphi1993}
{Sigurdsson}, S., \& {Phinney}, E.~S. 1993, \apj, 415, 631,
  \dodoi{10.1086/173190}

\bibitem[{{Sigurdsson} \& {Phinney}(1995)}]{Sigurdsson1995}
---. 1995, \apjs, 99, 609, \dodoi{10.1086/192199}

\bibitem[{{Spitzer}(1987)}]{Spitzer1987}
{Spitzer}, L. 1987, {Dynamical evolution of globular clusters}

\bibitem[{{Stephan} {et~al.}(2016){Stephan}, {Naoz}, {Ghez}, {Witzel},
  {Sitarski}, {Do}, \& {Kocsis}}]{step2016}
{Stephan}, A.~P., {Naoz}, S., {Ghez}, A.~M., {et~al.} 2016, \mnras, 460, 3494,
  \dodoi{10.1093/mnras/stw1220}

\bibitem[{{Stephan} {et~al.}(2019){Stephan}, {Naoz}, {Ghez}, {Morris},
  {Ciurlo}, {Do}, {Breivik}, {Coughlin}, \& {Rodriguez}}]{step2019}
---. 2019, \apj, 878, 58, \dodoi{10.3847/1538-4357/ab1e4d}

\bibitem[{{Tokovinin}(2014{\natexlab{a}})}]{tok14a}
{Tokovinin}, A. 2014{\natexlab{a}}, \aj, 147, 86,
  \dodoi{10.1088/0004-6256/147/4/86}

\bibitem[{{Tokovinin}(2014{\natexlab{b}})}]{tok14b}
---. 2014{\natexlab{b}}, \aj, 147, 87, \dodoi{10.1088/0004-6256/147/4/87}

\bibitem[{{Toonen} {et~al.}(2018){Toonen}, {Perets}, \& {Hamers}}]{toon2018}
{Toonen}, S., {Perets}, H.~B., \& {Hamers}, A.~S. 2018, \aap, 610, A22,
  \dodoi{10.1051/0004-6361/201731874}

\bibitem[{{Valtonen} \& {Karttunen}(2006)}]{valt2006}
{Valtonen}, M., \& {Karttunen}, H. 2006, {The Three-Body Problem}

\bibitem[{{Verbunt} {et~al.}(1984){Verbunt}, {van Paradijs}, \&
  {Elson}}]{Verbunt1984}
{Verbunt}, F., {van Paradijs}, J., \& {Elson}, R. 1984, \mnras, 210, 899,
  \dodoi{10.1093/mnras/210.4.899}

\bibitem[{{Vink} {et~al.}(2001){Vink}, {de Koter}, \& {Lamers}}]{Vink2001}
{Vink}, J.~S., {de Koter}, A., \& {Lamers}, H.~J.~G.~L.~M. 2001, \aap, 369,
  574, \dodoi{10.1051/0004-6361:20010127}

\bibitem[{{Ye} {et~al.}(2020){Ye}, {Fong}, {Kremer}, {Rodriguez}, {Chatterjee},
  {Fragione}, \& {Rasio}}]{ye2020}
{Ye}, C.~S., {Fong}, W.-f., {Kremer}, K., {et~al.} 2020, \apjl, 888, L10,
  \dodoi{10.3847/2041-8213/ab5dc5}

\bibitem[{{Ye} {et~al.}(2019){Ye}, {Kremer}, {Chatterjee}, {Rodriguez}, \&
  {Rasio}}]{Ye2019}
{Ye}, C.~S., {Kremer}, K., {Chatterjee}, S., {Rodriguez}, C.~L., \& {Rasio},
  F.~A. 2019, \apj, 877, 122, \dodoi{10.3847/1538-4357/ab1b21}

\bibitem[{{Zevin} {et~al.}(2018){Zevin}, {Samsing}, {Rodriguez}, {Haster}, \&
  {Ramirez-Ruiz}}]{Zevin2018}
{Zevin}, M., {Samsing}, J., {Rodriguez}, C., {Haster}, C.-J., \&
  {Ramirez-Ruiz}, E. 2018, ArXiv e-prints.
\newblock \doarXiv{1810.00901}

\end{thebibliography}

\end{document}